\documentclass[12pt]{iopart}

\expandafter\let\csname equation*\endcsname\relax
\expandafter\let\csname endequation*\endcsname\relax

\usepackage{epsfig}
\usepackage{epstopdf}
\usepackage{bm}
\usepackage{amsfonts}
\usepackage{amsmath}
\usepackage{amssymb}
\usepackage{mathrsfs}
\usepackage[lofdepth,lotdepth]{subfig}
\usepackage{xcolor, soul}

\def\d{\mathrm{d}}
\def\e{\mathrm{e}}

\begin{document}

\title[Uncertainty quantification in positron emission particle tracking]{A probabilistic framework for uncertainty quantification in positron emission particle tracking}

\author{Avshalom Offner}

\address{School of Mathematics and Maxwell Institute for Mathematical Sciences, The University of Edinburgh, Edinburgh EH9 3FD, UK}
\ead{avshalom.offner@ed.ac.uk}

\author{Sam Manger}

\address{Division of Cancer Sciences, Faculty of Biology, Medicine and Health, The University of Manchester, Manchester M13 9PL, UK}

\author{Jacques Vanneste}

\address{School of Mathematics and Maxwell Institute for Mathematical Sciences, The University of Edinburgh, Edinburgh EH9 3FD, UK}


\begin{abstract}
Positron Emission Particle Tracking (PEPT) is an imaging method for the visualization of fluid motion, capable of reconstructing three-dimensional trajectories of small tracer particles suspended in nearly any medium, including fluids that are opaque or contained within opaque vessels. The  particles are labeled radioactively, and their positions are reconstructed from the detection of pairs of back-to-back photons emitted by positron annihilation. Current reconstruction algorithms are heuristic and typically based on minimizing the distance between the particles and the so-called lines of response (LoRs) joining the detection points, while accounting for spurious LoRs generated by scattering. Here we develop a probabilistic framework for the Bayesian inference and uncertainty quantification of particle positions from PEPT data. We formulate a likelihood by describing the  emission of photons and their noisy detection as a Poisson process in the space of LoRs. We derive formulas for the corresponding Poisson rate in the case of cylindrical detectors, accounting for both undetected and scattered photons. We illustrate the formulation by quantifying the uncertainty in the reconstruction of the position of a single particle on a circular path from data generated by state-of-the-art Monte Carlo simulations. The results show how the observation time $\Delta t$ can be chosen optimally to balance the need for a large number of LoRs with the requirement of small particle displacement imposed by the assumption that the particle is static over $\Delta t$. We further show how  this assumption can be relaxed by inferring jointly the position and velocity of the particle, with clear benefits for the accuracy of the reconstruction.
\end{abstract}

%
%
%
%
%

\section{Introduction}

Standard methods for fluid flow visualization, e.g.\ particle image velocimetry (PIV) and particle tracking velocimetry (PTV), use high-speed cameras to capture the motion of small tracer particles that follow the fluid motion. Although widely used, these methods have two main drawbacks: the inability to visualize opaque flows -- either flows of opaque fluids or flows within opaque environments -- and a limited ability to visualize three-dimensional flows. Inspired by the physics of Positron Emission Tomography (PET), a standard method for diagnosis in nuclear medicine, Parker \textit{et al}. \cite{Parker1993} developed  Positron Emission Particle Tracking (PEPT), in which radioactively labeled particles serve as tracers of fluid flow. The radioactive decay of these  particles  emits positrons that are annihilated on or near the particles, resulting in $\gamma$-radiation that is picked up by an array of sensors, allowing for the triangulation of each particle position. Because of their high energy, the $\gamma$ photons penetrate even very dense media, making it possible to visualize, for example, the flow of  liquid metals \cite{Dybalska2020} or of colloidal drugs in the bloodstream \cite{Oerlemans2010}. The triangulation process naturally recovers the full position of the particles, so the reconstructed  trajectories are inherently three dimensional.

\begin{figure}
    \centering
     \includegraphics[trim=2cm 2cm 2cm 3cm,clip,scale=0.5]{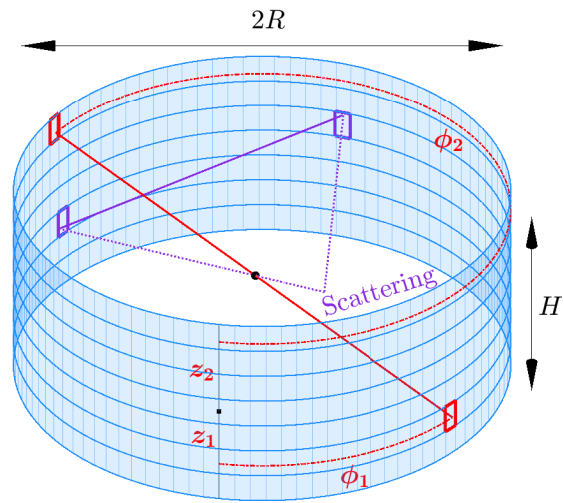}
    \caption{Schematic of a PEPT system: a representative positron-emitting particle (black) is positioned inside a cylindrical array of detectors with radius $R$ and height $H$. The highlighted cells denote photon detections and the solid lines represent Lines of Response (LoRs) that connect points of near-simultaneous detection. The red line represents a true back-to-back photon pair detection, indicative of the particle's true position; the purple line is an outlier LoR that results from a scattering event. The purple dotted line shows the true photon paths. The LoR parameterization $u=\left(\phi_{1},\phi_{2},z_{1},z_{2}\right)$, with $(z_i, \phi_i),\, i=1,2,$ the height and azimuthal angle detected photon $i$, is also shown.}
    \label{fig:System}
\end{figure}

The fundamental idea behind the triangulation of particle positions in PEPT is simple. In most annihilation events, the positron and electron masses emit two photons that travel in opposite directions to conserve momentum. The emitted photons are picked up at the detection surface (an array of small detectors, see figure \ref{fig:System}); two photons detected within a very short time interval are interpreted as having been generated by a single annihilation event. Such a pair detection represents a Line of Response (LoR) -- the straight line connecting  the two detection points -- along which the particle is presumably located (red line and black dot in figure \ref{fig:System}). 
Two nearly intersecting LoRs are seemingly sufficient to determine a particle instantaneous position with great accuracy. However, in practice, a much larger number of LoRs is required to cope with various sources of uncertainty. These include  Compton scattering \cite{Compton1923}, which leads to outlier LoR detections (illustrated by the purple lines in figure \ref{fig:System}), the finite size of the detectors, which limits resolution, and positron range, that is, the small but non-zero distance between the radioactive particle and the point of annihilation. The interested reader will find explanations on these and other sources of uncertainty in the paper by Moses \cite{Moses2011}. 

There is a rapidly growing body of work devoted to PEPT (see the recent book by Windows-Yule \textit{et al}. \cite{PEPTbook2022} and references therein). This includes the development of a variety of methods for the inversion of PEPT data \cite{Bickell2012,Blakemore2019,Manger2021,Nicusan2020,Odo2019,Wiggins2016,Wiggins2017}, which are assessed and compared in the recent review \cite{Windows-Yule2022}. The methods proposed to date are largely heuristic, and they do not estimate the error made on the recovered particle positions. The main aim of this paper is to remedy this and propose a systematic probabilistic framework for PEPT. This makes it possible to infer the particle positions and quantify their uncertainty using Bayesian methods \cite{GUM1995}.

We first formulate the forward problem, describing the probability of observing a set of LoRs given the positions and activities of the particles as a Poisson process in the space of LoRs and in time (\S\ref{sec:formulation}). We then give an expression for the corresponding rate in the case of a cylindrical detector, the most common geometry of PET/PEPT scanners, accounting for the detection of outlier LoRs resulting from scattering (\S\ref{sec:rate}).  We illustrate the benefits of our framework by carrying out Bayesian inference for synthetic data of a particle moving along a circular path (\S\ref{sec:application}). This provides a quantification of the uncertainty as a function of the time interval $\Delta t$ over which LoRs are observed, a key parameter that is determined by trial and error in other PEPT reconstruction algorithms. We also show how knowledge of the exact times of LoR detection can be leveraged to improve the accuracy of the inference.

\section{Formulation} \label{sec:formulation}
We consider $K$ radioactive particles with positions $\bm{X}(t)=\left\{\bm{x}_{1}\left(t\right),\cdots,\bm{x}_{K}\left(t\right)\right\}$, suspended in a fluid within the volume enclosed by an  array of detectors (see figure \ref{fig:System} for a schematic of a system with a cylindrical array). The particles spontaneously emit positrons at random times, according to independent Poisson processes whose rates are the activities $\varrho=\left\{\rho_{1},\cdots,\rho_{K}\right\}$. These activities can be taken as constant over the time $\Delta t$ of observation of a batch of LoRs. Positrons released from radioactive decay are annihilated close to the particles, emitting a pair of back-to-back photons in a random direction. The photons are subsequently picked up by a pair of detectors. The LoRs joining each pairs of detectors constitute the data from which the particle positions $\bm{X}(t)$ should to be retrieved. We adopt a Bayesian approach and formulate a forward model, which gives the probability of observing a set of LoRs given the particle positions $\bm{X}(t)$ and activities $\varrho$ and is translated into a probability for $\bm{X}(t)$ and $\varrho$ using Bayes's formula. The forward model is best expressed as a Poisson process in the space of LoRs.

\subsection{Forward model}

\begin{figure}
    \centering
    \includegraphics[trim=2cm 7cm 2.5cm 3cm,clip,scale=0.6]{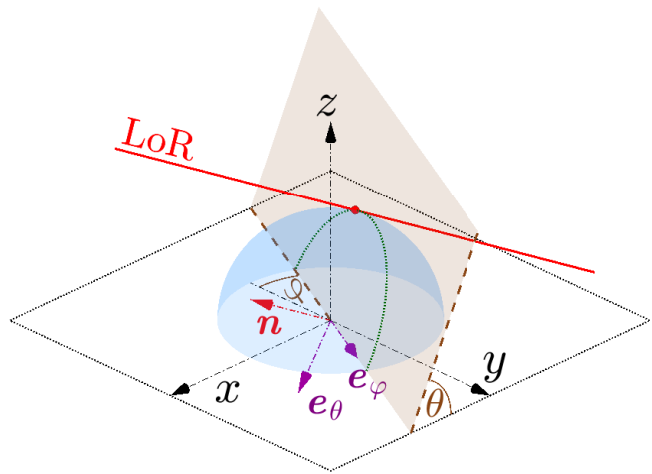}
    \caption{Parameterization of LoRs: the direction $\bm{n}$ of a LoR is given by the polar spherical angles $\varphi$ and $\theta$. The brown plane is normal to $\bm{n}$ and goes through the origin; its intersection with the hemisphere tangent to the LoR is shown as a green arc. The point on the LoR closest to the origin (red dot) is parameterized by the two coordinates $(a_{\varphi},a_{\theta})$ in the planar frame $(\bm{e}_{\varphi}, \bm{e}_{\theta})$, with $(\bm{n},\bm{e}_{\theta},\bm{e}_\varphi)$ the standard unit vectors of spherical polar coordinates. }
    \label{fig:Schematics}
\end{figure}

The space of lines in $\mathbb{R}^3$ is 4-dimensional since lines can be characterized by a unit vector $\bm{n}$, giving their direction, and the closest point to the origin, which lies in the plane normal to $\bm{n}$. With the restriction $n_z \ge 0$, this identifies the space of lines with the tangent bundle to a hemisphere. We parameterize $\bm{n}$ using the standard polar spherical angles. The point on the line closest to the origin is then conveniently represented as 
\begin{equation}
\bm{a}=a_{\varphi}\bm{e}_{\varphi}+a_{\theta}\bm{e}_{\theta}
\end{equation}
in terms of the unit vectors $\bm{e}_{\varphi}$ and $\bm{e}_{\theta}$ of the spherical coordinate system (with $\bm{e}_r = \bm{n}$, see figure \ref{fig:Schematics}). In this way, 
a LoR $L$ is parameterized as 
\begin{equation}
L=\left(\varphi,\theta,a_{\varphi},a_{\theta}\right)\in\left[0,2\pi\right]\times\left[0,\pi/2\right]\times\mathbb{R}^{2}. \label{eq:parameterization}
\end{equation}
There is a natural measure on that space, $\d \mu$ say, characterized by invariance in rotation and translation \cite{Beckers1990}. In terms of the coordinates \eqref{eq:parameterization}, this has the simple form
\begin{equation}
\d \mu = \sin \theta \, \d \varphi \d \theta \d a_\varphi \d a_\theta.
\label{dmu}
\end{equation}

The generation of LoRs given the positions and activities of $K$ radioactive particles defines a Poisson point process in the space of lines and in time. This has long been known in PET, with numerous works describing LoR generation as a Poisson process (see, e.g., \cite{Shepp1982,Parra1998}), however in PEPT the radioactive source position is inherently time dependent. We denote by $\lambda(L|\bm{x}(t),\rho)$ the rate (or intensity) of the Poisson process corresponding to the generation of LoRs by a single particle with activity $\rho$ moving along the path $\bm{x}(t)$. Thus $\lambda \d \mu \d t$ is the probability of finding a LoR in the volume $\left[\varphi,\varphi+\d\varphi\right]\times\left[\theta,\theta+\d\theta\right]\times\left[a_{\varphi},a_{\varphi}+\d a_{\varphi}\right]\times\left[a_{\theta},a_{\theta}+\d a_{\theta}\right]$ during a time interval $\left[t,t+\d t\right]$. The form of $\lambda(L|\bm{x}(\cdot),\rho)$ depends on the geometry of the detectors and physics of positron annihilation, scattering and other processes; we give an explicit form for a cylindrical detector in \S\ref{sec:rate}. Note that $\lambda(L|\bm{x}(t),\rho)$ depends on time implicitly through the time dependence of the particle position $\bm{x}$.
The rate of generation by $K$ particles is simply $\sum_{k=1}^K$ $\lambda(L|\bm{x}_k(t),\rho_k)$.

Using the standard formula for Poisson point processes \cite{streit2010poisson}, we can write down the probability of observing $N$ LoRs $L_1,\cdots,L_N$ at times $0<t_1<\cdots<t_N<\Delta t$
as
\begin{equation}
P\left(\mathcal{L}_t|\bm{X}(\cdot),\varrho,\Delta t\right)=\e^{-\Lambda(\bm{X}(\cdot),\varrho,\Delta t)}\prod_{n=1}^N \sum_{k=1}^K \lambda\left(L_{n} |\bm{x}_k(t_n),\rho_k \right). \label{eq:probability PoissonTime}
\end{equation}
Here $\mathcal{L}_t=\left\{ (L_{1},t_1),\cdots,(L_{N},t_N)\right\}$ denotes the observed LoRs and times of observation and we have defined
\begin{equation}
\Lambda(\bm{X}(\cdot),\varrho,\Delta t)= \sum_{k=1}^K  \int_0^{\Delta t} \int   \lambda\left(\left.L\right|\bm{x}_k(t),\rho_k\right)\d\mu \d t, \label{eq:Lambda0}
\end{equation} 
with the integration with respect to $\d \mu$ carried out over the entire space of LoRs. Note that  $P\left(\mathcal{L}_t|\bm{X}(\cdot),\varrho,\Delta t \right)$ is properly normalized, in the sense that
\begin{equation}
\sum_{N=0}^\infty \int_0^{\Delta t} \d t_1 \int_{t_1}^{\Delta t} \d t_2 \cdots \int_{t_{N-1}}^{\Delta t} \d t_N \int\cdots\int P\left(\mathcal{L}_t|\bm{X}(\cdot),\varrho,\Delta t \right) \, \d \mu_1 \cdots \d \mu_N =1.    
\end{equation}

In most inversion algorithms, the time interval $\Delta t$ is taken small enough that particles can be assumed fixed (see Ref.\ \cite{Lee2015}, however). With this assumption, information about the times of observations of the LoRs can be disregarded, and we can focus on the probability of observing a set $\mathcal{L}=\{L_1,\cdots,L_N\}$ of LoRs, irrespective of the times. This probability is obtained by integrating \eqref{eq:probability PoissonTime} with respect to  $0< t_1 < t_2 < \cdots < t_N < \Delta t$ to find
\begin{equation}
P\left(\mathcal{L}|\bm{X},\varrho,\Delta t\right)=\frac{ (\Delta t)^N \e^{-\Lambda(\bm{X},\varrho,\Delta t)}}{N!} \prod_{n=1}^N \sum_{k=1}^K \lambda\left(L_{n} |\bm{x}_k,\rho_k \right), \label{eq:probability Poisson}
\end{equation}
where the notation emphasizes that $\bm{X}$ and $\bm{x}_k$ are regarded as constant.
Eq.\ \eqref{eq:Lambda0} also simplifies to
\begin{equation}
\Lambda(\bm{X},\varrho,\Delta t)=\Delta t \sum_{k=1}^K \int \lambda\left(\left.L\right|\bm{x}_k,\rho_k\right)\d\mu. \label{eq:Lambda}
\end{equation}

\subsection{Bayesian inference} \label{sec:bayesian}

Bayes' formula provides the means to infer a probability distribution for the particle positions and activities from LoR data. Under the assumption of fixed particles, it takes the form
\begin{equation}
P(\bm{X},\varrho |\mathcal{L},\Delta t) \propto P(\mathcal{L}|\bm{X},\varrho,\Delta t)P(\bm{X},\varrho), \label{eq:Bayes}
\end{equation}
where $P(\bm{X},\varrho |\mathcal{L},\Delta t)$ is the posterior probability for  the (fixed) particle positions $\bm{X}=\left\{\bm{x}_1,\cdots,\bm{x}_K\right\}$ and activities $\varrho=\left\{\rho_1,\cdots,\rho_K\right\}$, $P(\bm{X},\varrho |\mathcal{L},\Delta t)$ is the likelihood given in \eqref{eq:probability Poisson}, and $P(\bm{X},\varrho)$ is the prior.
In the application of \S\ref{sec:application}, we take a uniform (improper) prior over the set of physically realisable particle positions and activities, namely
\begin{equation}
P\left(\bm{X},\varrho\right)=
\begin{cases}
1 & \textrm{if} \ x_{k}\in V \ \textrm{and} \ \rho_{k}\geq0 \quad \forall k \\
0& \textrm{otherwise}
\end{cases},
\label{eq:prior}
\end{equation}
where $V$ is the volume enclosed by the detectors.

The assumption of fixed  particle positions that underlies \eqref{eq:Bayes} restricts the length of the observation time $\Delta t$ and hence of the number of observed LoRs. Depending on the speed of the particles, this can severely limit the accuracy of the inversion.  
To overcome this, we can retain information about the times of LoR observations and carry out Bayesian inference for the particle trajectories rather than fixed positions, using \eqref{eq:probability PoissonTime} as the likelihood. In practice, the trajectories must be parameterized, say as $\bm{x}_k(t)=\hat{\bm{x}}(t,\vartheta_k)$, where $\vartheta_k$ groups the parameters of the trajectory of particle $k$. Bayes' formula is then used in the form
\begin{equation}
 P(\bm{\vartheta},\varrho |\mathcal{L}_t,\Delta t) \propto P(\mathcal{L}_t|\bm{\vartheta},\varrho,\Delta t)P(\bm{\vartheta},\varrho), \label{eq:Bayes_t}   
\end{equation}
where $\bm{\vartheta}=\{\vartheta_1,\cdots,\vartheta_k\}$ and the likelihood $P(\mathcal{L}_t|\bm{\vartheta},\varrho,\Delta t)$ is obtained from \eqref{eq:probability PoissonTime} by substituting the parametric form of the trajectories. 
A simple parameterization can be constructed by a Taylor expansion of the particle trajectory around the middle of the observation time. We  follow Manger \textit{et al.} \cite{Manger2021} by writing
\begin{equation}
\bm{x}_k\left(t\right)=\sum_{m=0}^{M} \frac{1}{m!} \left.\frac{\mathrm{d}^{m}\bm{x}_k}{\mathrm{d}t^{m}}\right|_{\Delta t/2}\left(t-\Delta t/2\right)^{m}+O\left(\Delta t^{M+1}\right). \label{eq:Taylor series}
\end{equation}
and use the position and its $M$ first derivatives as parameters:
\begin{equation}
   \vartheta_k=\left( \left. \bm{x}_k \right|_{\Delta t/2},\cdots,\left. \frac{\d^M\bm{x}_k}{\d t^M} \right|_{\Delta t/2} \right).
\end{equation}
We refer to inference of this type as high-order inference and to $M$ as the order of the inference. 
We implement the $M=1$, first-order inference in \S\ref{sec:high-order inference}. 

\section{Poisson rate} \label{sec:rate}  

We now derive the Poisson rate  $\lambda(L|\bm{x},\rho)$ for a cylindrical detector configuration  with radius $R$ and height $H$  as depicted in figure \ref{fig:System}. We assume that positron annihilation takes place close enough to the radioactive particle that the photons can be assumed to travel on a line through $\bm{x}$. This  line has a random direction, distributed uniformly and parameterized by two spherical angles $\varphi'$ and $\theta'$, say, with probability density
\begin{equation}
    p(\varphi',\theta') = \frac{\sin \theta'}{2 \pi}.
    \label{uniangle}
\end{equation}
The angles $\varphi'$ and $\theta'$ differ from the angles $\varphi$ and $\theta$ attributed to the LoR because of measurement errors which we now model.

\subsection{Detection error} \label{sec:detection error}
The LoRs are detected on a surface of discrete detectors, in our case a cylindrical envelope. The line through $\bm{x}$ intersects this cylinder at two points which we parameterize as 
\begin{equation}
u'=\left(\phi'_{1},\phi'_{2},z'_{1},z'_{2}\right)\in\left[0,2\pi\right]\times\left[0,2\pi\right]\times\left[-H/2,H/2\right]\times\left[-H/2,H/2\right], \label{eq:u'}
\end{equation}
where $\phi'_{1,2}$ and $z'_{1,2}$ are the azimuthal angles and heights of the two points (see figure \ref{fig:System} for illustration of this parameterization). Measurement uncertainty introduces a deviation between the detected coordinates $u$ of these intersections and the exact intersection $u'$. As a result, the coordinates $L=(\varphi,\theta,a_\varphi,a_\theta)$ of the observed LoR  differ from the coordinates $L'=(\varphi',\theta',x_{\varphi'},x_{\theta'})$ of the line followed by the photons. The uncertainty associated with the finite size of the detectors implies that the intersection parameters $u$ of the observed LoR are distributed uniformly over the area of the two detectors intersected by the true LoR. To avoid discontinuous distribution and to group all measurement errors in a single, simple model, we replace this uniform distribution by a normal distribution: we assume that $u$ is normally distributed about $u'$ according to 
\begin{equation}
u | L' \sim \mathcal{N}(u',C), \quad 
\textrm{where} \ \ C=\left(\begin{array}{cccc}
g^{2}/R^{2} & 0 & 0 & 0\\
0 & g^{2}/R^{2} & 0 & 0\\
0 & 0 & g^{2} & 0\\
0 & 0 & 0 & g^{2}
\end{array}\right).
\label{unormal}
\end{equation}
Here $g^{2}$ is the variance of the detected heights $z_1$ and $z_2$; assuming that the uncertainty about the true LoR is isotropic, it follows that the variance in $\phi_{1},\phi_{2}$ is $g^2/R^2$. In practice, we fix $g$ to ensure the Gaussian distribution does not strongly deviate from a top hat function; specifically, we take $g$ such that the difference between integrating the top hat and Gaussian distributions over a single detector is minimized.


There is a one-to-one map between the coordinates of LoRs and the coordinates $u$ of the detected points which we denote by $u=F\left(L\right)$ (see appendix \ref{ap:mapping} for its explicit form). In the linear approximation, the normal distribution \eqref{unormal} implies that $L$ is also normal, specifically
\begin{equation}
    L | L' \sim \mathcal{N}(L',\Sigma), \quad \textrm{where} \ \ \Sigma=\left[\nabla F\left(L\right)\right]^{-1}C \left[\nabla F\left(L\right)\right]^{-\mathrm{T}}.
    \label{L|L'}
\end{equation}

We deduce the probability that a LoR $L$ is detected given that a positron is annihilated at $\bm{x}$ 
by marginalising over the angles $\phi'$ and $\theta'$ of $L'$ distributed according to \eqref{uniangle}. Using that $g/R \ll 1$, we can approximate the Gaussian in $\phi'$ and $\theta'$ by $\delta(\phi'-\phi)\delta(\theta'-\theta)$ to obtain this probability as
\begin{equation}
    \frac{\sin \theta}{2\pi\left|\det\Sigma_{2}\right|^{1/2}}\exp\left(-\bm{\Delta}^{T}\Sigma_{2}^{-1}\bm{\Delta}/2\right),
\end{equation}
where $\Sigma_{2}$ is the lower-right $2\times2$ block of $\Sigma$, given explicitly in \eqref{eq:Sigma2}, and
\begin{equation}
\bm{\Delta}=\left(\begin{array}{c}
\Delta_{\varphi}\\
\Delta_{\theta}
\end{array}\right)=\left(\begin{array}{c}
a_{\varphi}-x_{\varphi}\\
a_{\theta}-x_{\theta}
\end{array}\right).
\end{equation}
Finally, the intensity of the Poisson process in the space of LoRs and time is obtained as

\begin{equation}
\lambda\left(\left.L\right|\bm{x},\rho\right)=\frac{\rho}{2\pi\left|\det\Sigma_{2}\right|^{1/2}}\exp\left(-\bm{\Delta}^{T}\Sigma_{2}^{-1}\bm{\Delta}/2\right) \, \chi(L \in \mathcal{D}(\bm{x}_k))
, \label{eq:intensity single}
\end{equation}
with the factor $\sin \theta$ absorbed in the volume element $\d \mu$ in \eqref{dmu}. 
The characteristic function $\chi(\cdot)$ in \eqref{eq:intensity single} ensures that $\lambda$ is non-zero only if $L$ belongs to the set $\mathcal{D}(\bm{x}_k)$ of LoRs through $\bm{x}_k$ that intersect the detector.

A particularly simple expression for $\lambda$ is obtained under the assumption $a_{\varphi},a_{\theta}\ll R$ valid when the particle is located well away from the detector, as is the case for experiments carried out in small vessels.  
Eq.\ \eqref{eq:intensity single} is then well approximated by setting $a_\varphi=a_\theta=0$ to obtain
\begin{equation}
\lambda\left(\left.L\right|\bm{x},\rho\right)=\frac{\rho}{\pi g^{2}\sin\theta}\exp\left(-\left[\frac{\Delta_{\varphi}^{2}}{g^2}+\frac{\Delta_{\theta}^{2}}{g^{2}\sin^2 \theta}\right]\right) \, \chi(L \in \mathcal{D}(\bm{x}_k)).
\end{equation}

\subsection{Outliers}
Statistically, a proportion of the photons emitted from annihilation events are scattered, resulting in outlier LoR detections (see figure \ref{fig:System} for a visualization of such detection). We account for these outliers by adding to the Poisson process describing the generation of LoRs by $K$ particles a component independent of the particle positions and activities.
This captures the fact that scattering leads to the identification of spurious LoRs whose distribution is not directly related to the particle positions.
For simplicity, we assume this distribution to be uniform over the entire space of LoRs. We therefore re-write the probability density \eqref{eq:probability Poisson} for the detected LoRs as
 \begin{equation}
P\left(\mathcal{L}|\bm{X},\varrho,\Delta t\right)=\frac{(\Delta t)^N \e^{-\Lambda(\bm{X},\varrho,\Delta t)}}{N!} \prod_{n=1}^N \left( \frac{\rho_0}{ \sigma_0^2} + \sum_{k=1}^K \lambda\left(L_{n} |\bm{x}_k,\rho_k \right) \right), \label{eq:probabilityfull}
\end{equation}
where
\begin{equation}
\Lambda\left(\bm{X},\varrho,\Delta t\right)=\Delta t\left(\rho_{0}+\sum_{k=1}^{K}\int_{\mathcal{D}}\lambda\left(\left.L\right|\bm{x}_{k},\rho_{k}\right)\,\d \mu\right),
    \label{eq:Lambda full}
\end{equation}
in which $\mathcal{D}$ is the set of all the possible LoRs for a specific detector configuration, $\rho_0$ is the rate of scattered events, which is treated as independent of the activities $\rho_k$ for simplicity, and
\begin{equation}
\sigma_{0}^{2}=\int_{\mathcal{D}}\d\mu \label{eq:sigma0 def}
\end{equation}
is the outlier variance. Note that the activities $\rho_{k}$ now stand for the rate of generation of `true' LoRs, as opposed to outlier LoR resulting from scattering events. The total activity from all particles is $\sum_{k=0}^{K}\rho_{k}$. 

The evaluation of the integrals \eqref{eq:Lambda full} and \eqref{eq:sigma0 def} defining  $\Lambda$ and $\sigma_0$ requires to determine the sets $\mathcal{D}$ and $\mathcal{D}(\bm{x_k})$ of, respectively, all LoRs intersecting the detectors twice and  LoRs through $\bm{x}_k$ intersecting the detectors twice. We detail in Appendices \ref{ap:sigma0} and \ref{ap:G} the parameterization of these sets and the computation of the integrals. The outlier variance is found as
\begin{equation}
\sigma_{0}^{2}=\frac{\pi HR}{8\beta}\left(2\beta+1-\sqrt{4\beta^{2}+1}\right), \label{eq:sigma0^2}
\end{equation}
where $\beta=R/H$. This is proportional to the detecting surface area $A_{d}=2\pi RH$. In the limit $\beta\rightarrow0$ corresponding to a cylindrical detector that resembles a long tube, all LoRs are detected and $\sigma_{0}^2\rightarrow A_{d}/8$. In the limit $\beta\rightarrow\infty$, corresponding to a detecting surface that resembles a ring in which no true LoRs are detected, $\sigma_{0}^2\rightarrow0$ as expected. The rate $\Lambda$ is approximated as
\begin{equation}
\Lambda\left(\bm{X},\varrho,\Delta t\right)=\Delta t\left(\rho_{0}+\sum_{k=1}^{K}\rho_{k}\mathcal{G}\left(\bm{x}_{k}\right)\right),
\label{eq:Lambda2}
\end{equation}
assuming that the distance between the particles and the detectors in much larger than $g$. Since $g \ll R$, this assumption holds for realistic PEPT setups. Here, $\mathcal{G}\left(\bm{x}\right)$ is a solid angle subtended by the cylindrical detector from $\bm{x}_k$ divided by $4\pi$; it can be written as
\begin{equation}
\mathcal{G}\left(\bm{x}\right)=\frac{1}{2\pi}\int_{0}^{2\pi}\d\varphi\int_{\theta_{\text{min}}}^{\pi/2}\sin\theta\,\d\theta=\frac{1}{2\pi}\int_{0}^{2\pi}\cos\theta_{\text{min}}\left(\bm{x},\varphi\right)\,\d\varphi, \label{eq:G_k def}
\end{equation}
where $\theta_\mathrm{min}$, given explicitly in \eqref{eq:thetamin}, is the minimum angle required for the LoR to intersect the cylindrical detector twice. 

\begin{figure}
    \centering
    \includegraphics[trim=3cm 2cm 3cm 1cm,clip,scale=0.35]{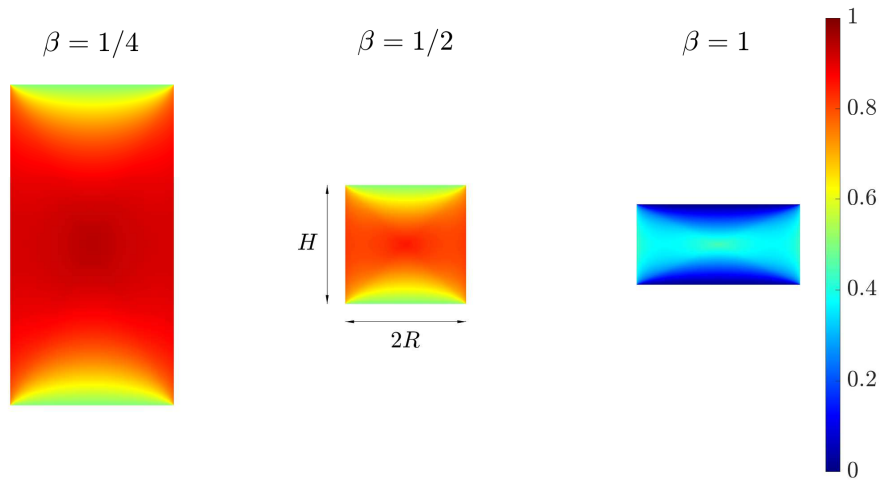}
    \caption{Function $\mathcal{G}\left(r,z\right)$, computed numerically from \eqref{eq:Gk appendix} for the aspect ratio $\beta=R/H=1/4$ (left), $1/2$ (center), and $1$ (right). The function quantifies the reduction in particle detectability depending on its position relative to the detector.}
    \label{fig:G rz}
\end{figure}

According to \eqref{eq:Lambda2}, $\rho_k \mathcal{G}(\bm{x}_k)$ can be interpreted as an `effective' activity for particle $k$ that accounts for failure of detection of LoRs caused by detector geometry through the factor $0\le \mathcal{G}(\bm{x}_k) \le 1$.
Figure \ref{fig:G rz} shows $\mathcal{G}$ as a function of the cylindrical coordinates $(r,z)$ of the particle ($\mathcal{G}$ is independent of the azimuthal angle by symmetry) for three aspect ratios $\beta=R/H$.
As $\beta$ is increased $\mathcal{G}$ trivially decreases throughout the domain; the maximum value is always obtained at the origin $r=z=0$. We derive asymptotic approximations for $\mathcal{G}$ at small and large $\beta$ in appendix \ref{ap:G}.

\section{Application} \label{sec:application}

To demonstrate the effectiveness of our framework, we carry out the Bayesian inference of the instantaneous position and activity in a simulation of a single radioactive particle moving along a circle. The circle is concentric with the cylindrical detector surface so that, by symmetry, $\mathcal{G}(\bm{x})$ is  constant along the particle trajectory. Extending the simulation and Bayesian inference to multiple particles is straightforward in principle; however, the corresponding increase in the number of inferred parameters complicates the analysis and significantly increases the computational cost. Considering an upper limit of a few hundred parameters in Bayesian inference \cite{BayesianDataAnalysis2013}, a realistic limit on the number of particles and/or the inference order can be easily estimated..

\subsection{Simulations}

We carry out our analysis on synthetic data produced by GATE \cite{Jan2004}, a unique software for positron emission simulations. This has the benefit of providing the exact particle position as ground truth, thus enabling the exact measurement of the inference error. We consider a cylindrical detector surface with $R=200$ and $H=230\,\mathrm{mm}$ ($\beta=R/H=0.87$), composed of identical $4 \times 4$ mm detector elements. The rotating, positron-emitting particle is suspended in water, and its `corrected' activity, that is accounting for the effect of detector deadtime which is not represented explicitly in the present work, is $\rho=5\cdot10^4$ emissions/s. LoRs are recorded over a period of one second, providing ample data for the inference. We carry out simulations in four different conditions with varying circle radii and rotation frequencies: $r=50$ mm at $f=1$ and 2 Hz, and $r=100$ mm at $f=0.5$ and 1 Hz. This yields two distinct particle velocities $2\pi f r\approx0.3$ and 0.6 m/s, each repeated twice at varying acceleration $4\pi^2 f^2 r$. 

\subsection{Data analysis}

We employ a standard Metropolis--Hastings  Monte Carlo Markov Chain (MCMC) algorithm \cite{Hastings1970} to sample the posterior distribution for the particle position $\bm{x}$ and activities $\rho_0$ and $\rho_1$. The likelihood and prior are those given in \eqref{eq:probabilityfull}, with $\lambda$ in \eqref{eq:intensity single}, and \eqref{eq:prior}. For the current setup with $4 \times 4$ mm$^2$ detectors, the detecting error variance in \eqref{eq:intensity single} gives $g=2.43$ mm.

The MCMC algorithm produces samples from the posterior distribution as a sequence of (statistically dependent) sets of inferred parameters. This sequence is constructed iteratively, with a new set of parameters obtained by proposing a random change to a single randomly chosen parameter, then accepting or rejecting the change depending on a likelihood ratio \cite{BayesianDataAnalysis2013}. We use Gaussians as proposal density functions for the difference between each old and new parameter. We increase/decrease the variance of these Gaussians by a factor of two if 40 consecutive steps were accepted/rejected, which resulted in an averaged acceptance rate of $0.26$. The initial values of the Gaussian proposal density are set as follows: the position variance is initially set to $g^2$; the activity standard deviation is initially taken as 5\% of the particle activity. We use $10^5$ steps to infer the position and activity at each run. We start each MCMC sampling with $\rho_{0}=\rho_{1}=\rho/2$ (with $\rho$ the prescribed particle activity); to speed up the computation, we start from the exact particle position thus not requiring a burn-in period for the MCMC.

For each pair of values of the radius $r$ and frequency $f$ of the particle trajectory, we infer the position and activity of the particle at 10 equally spaced times $t_i$ using the LoRs collected in the time interval $t_{i}-\Delta t/2<t_{i}<t_{i}+\Delta t/2$. The statistics of each set of 10 MCMC samplings makes it possible to characterize the variability of the inference process that arises from the randomness of the positron emission. Our main focus is on the dependence of the accuracy of the inference on the choice of the observation interval $\Delta t$. We therefore run the entire MCMC computation for a range of values of $\Delta t$ and compare the results. 

\subsection{Results} \label{sec:results}

We first show the results of a single MCMC sampling, carried out for a particle rotating with frequency $f=1\,\mathrm{Hz}$ at a radius $r=50\,\mathrm{mm}$, resulting in the tangential velocity $v=2\pi f r\approx0.3\:\mathrm{m/s}$. The time interval is chosen as $\Delta t=10\,\mathrm{ms}$ which yields approximately 250 detected LoRs. 

\begin{figure}
\centering
\subfloat[][]{\includegraphics[scale=0.27]{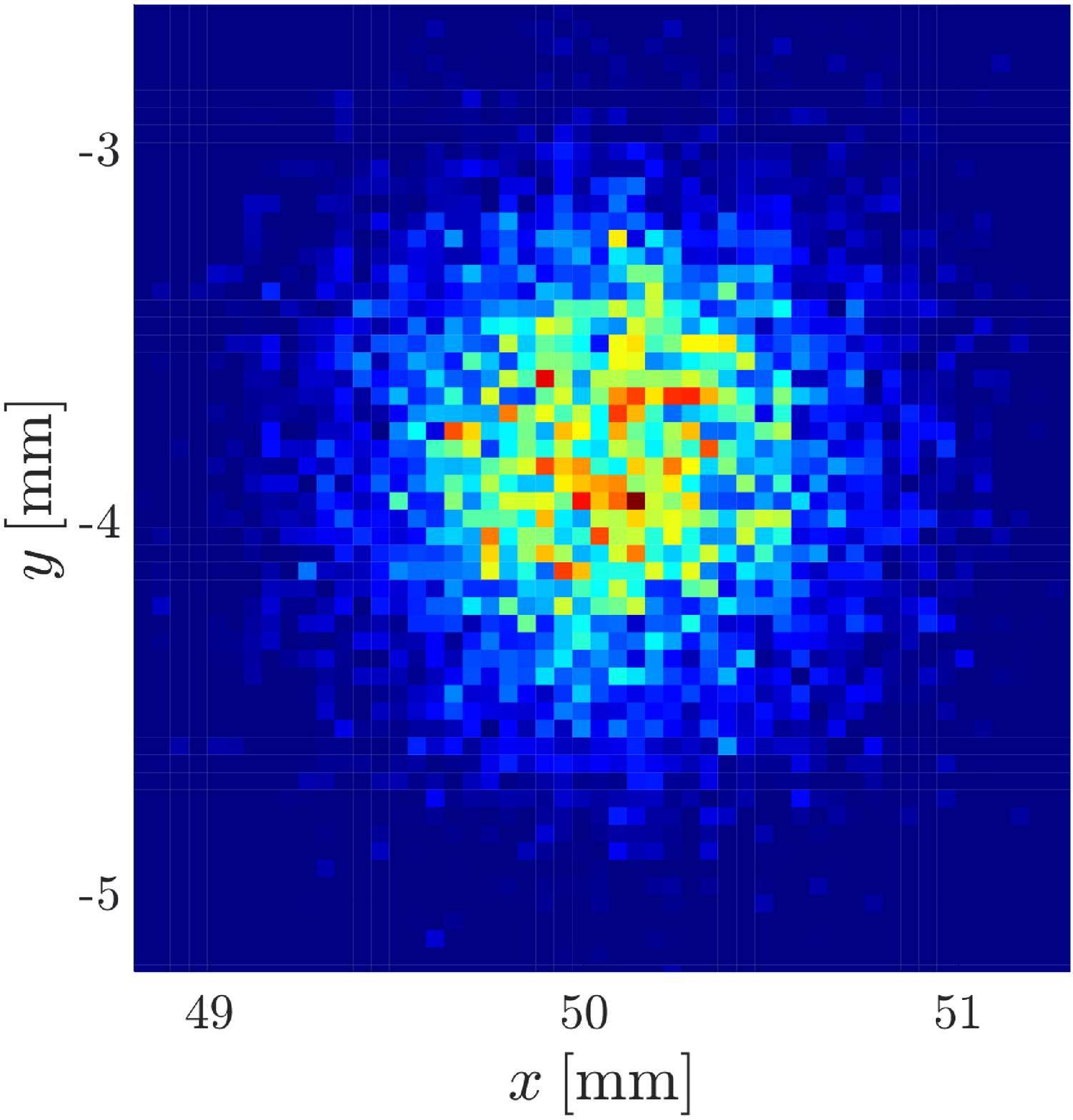}
\label{fig:x-y pdf}}
\qquad
\subfloat[][]{\includegraphics[scale=0.27]{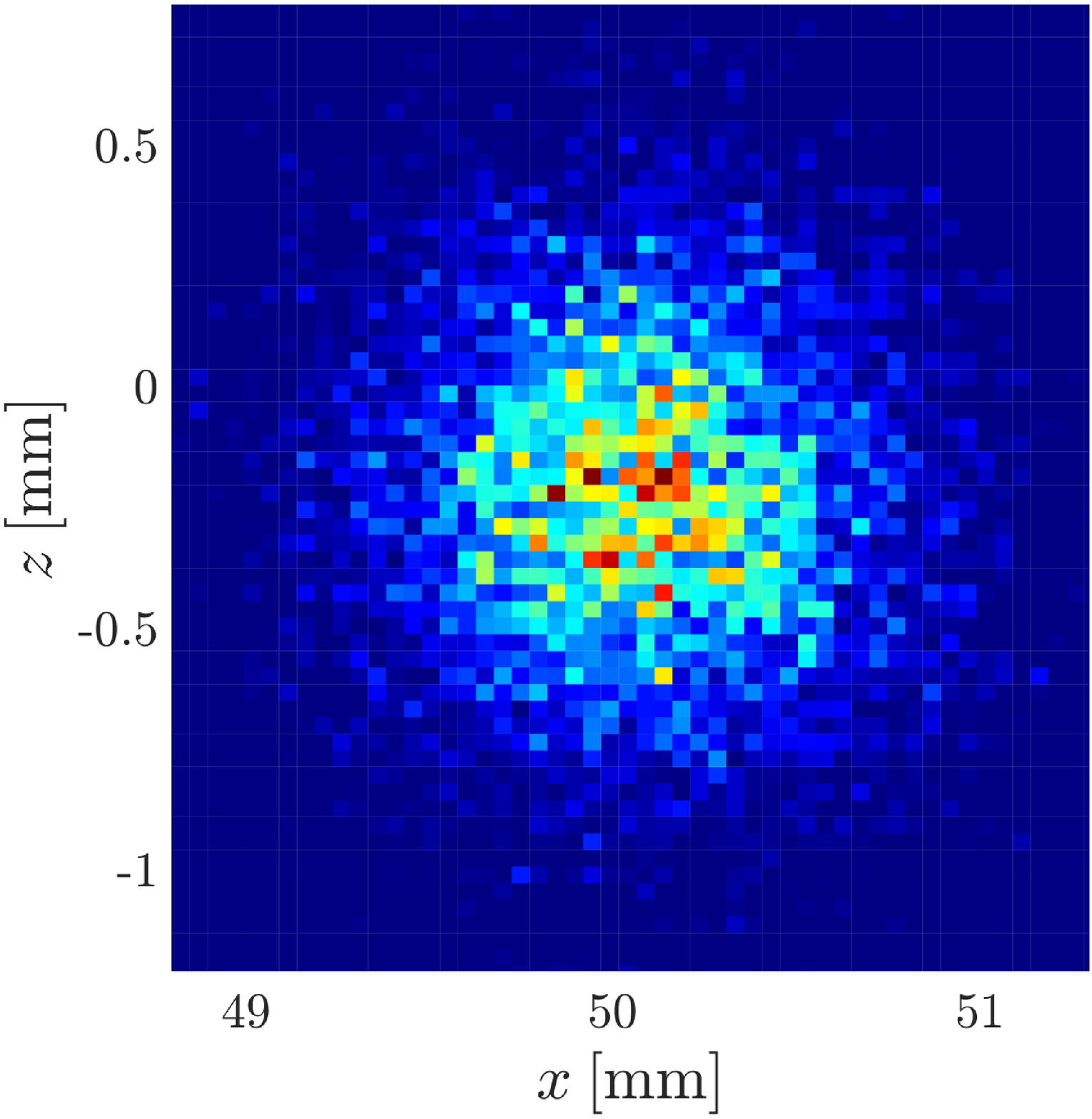}
\label{fig:x-z pdf}}

\caption{Posterior pdf for the instantaneous position of a particle moving at $v=0.3\:\mathrm{m/s}$ estimated by MCMC sampling using \eqref{eq:probabilityfull} with $\Delta t=10\:\mathrm{ms}$. The figures show conditional pdfs in the (a) horizontal plane $(x,y,0)$, and (b) vertical plane $(x,-3.56,z)$.
The majority of steps in the chain are enclosed within a 2 mm edge cube, demonstrating the localization of the PDF. For reference, the true position at the midst of the time interval is $\bm{x}=(49.87,-3.56,0)$ mm.}
\label{fig:pdfs}
\end{figure}

In figure \ref{fig:pdfs} we show two-dimensional histograms of the posterior pdf for the particle position in the $(x,y,0)$ horizontal (a) and $(x,-3.56,z)$ vertical (b) planes. (the particle true position at the midst of the time interval is $\bm{x}=\left(49.87,-3.56,0\right)$) The results illustrate the localization of the posterior defined by \eqref{eq:probabilityfull}, with nearly the entire pdf supported in a $2\times 2 \, \mathrm{mm}^2$ area in both the horizontal and vertical. To quantify the uncertainty in the inference of the particle position more systematically, 
we estimate a covariance matrix for the particle position from the MCMC samples. We use its three eigenvalues, $\mathscr{L}_i$ say, to compute the
metric
\begin{equation}
s=\mathcal{R}_{3}\left(\prod_{i=1}^{3}\sqrt{\mathscr{L}_{i}}\right)^{1/3}, \label{eq:s}
\end{equation}
where $\mathcal{R}_{3}=2.79$, such that $s$ is the radius of a sphere that marks the 95\% confidence level for the Gaussian distribution with the estimated covariance. ($\mathcal{R}_{3}$ is the three-dimensional equivalent of the more famous $\mathcal{R}_{1}=1.96$ that gives the 95\% confidence interval for a one-dimensional Gaussian distribution.) The pdf in figure \ref{fig:pdfs} has $s=0.79$ mm, five times smaller than the detector side of 4 mm. 

\begin{figure}
\centering
\subfloat[][]{\includegraphics[scale=0.27]{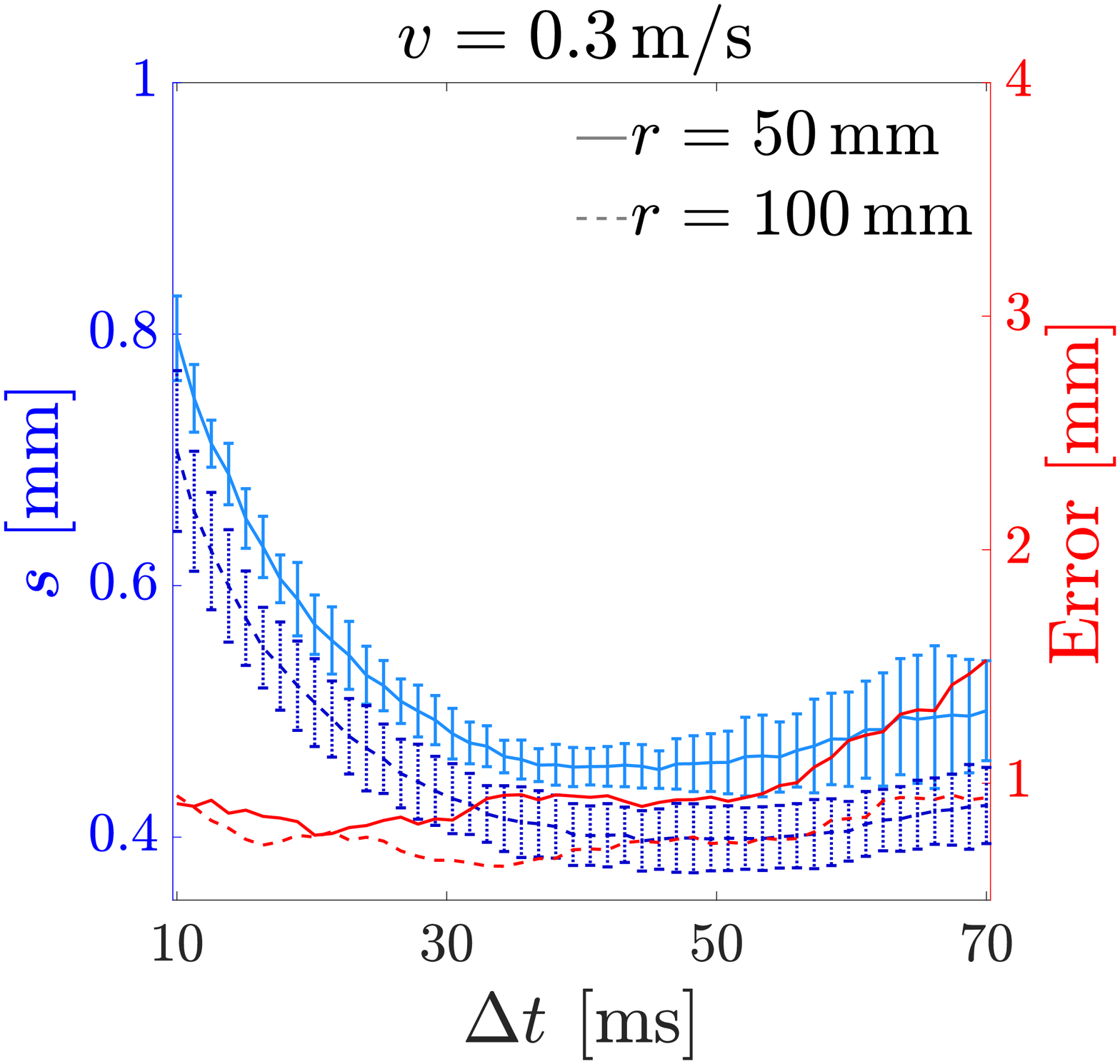}
\label{fig:s vs dt T0 v0.3}}
\qquad
\subfloat[][]{\includegraphics[scale=0.27]{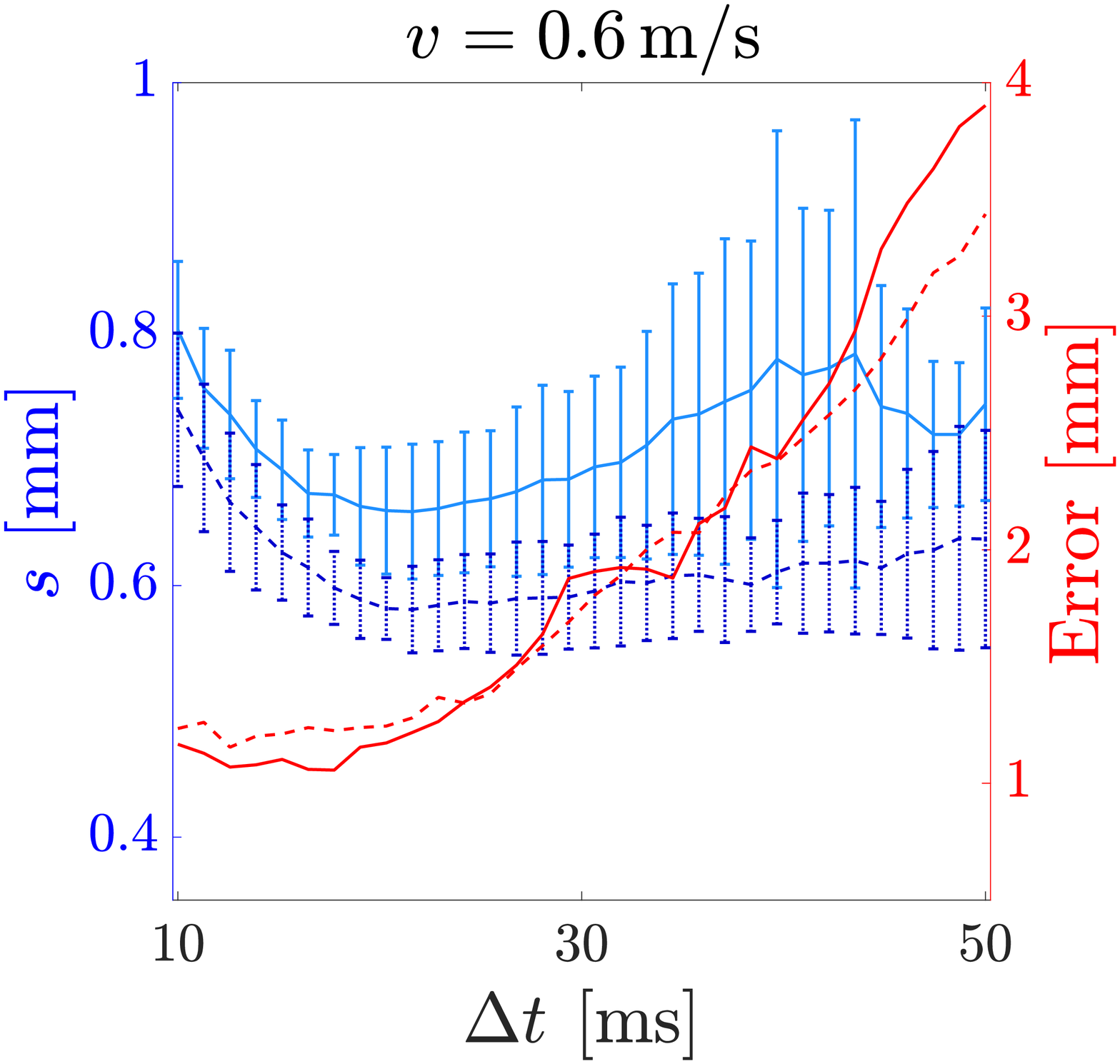}
\label{fig:s vs dt T0 v0.6}}

    \caption{Uncertainty metric $s$ defined in 
    \eqref{eq:s} (blue curves, left axes) and error $|\bm{x}-\bm{x_*}|$ in the inferred  instantaneous particle position (red curves, right axes) as functions of the observation time interval $\Delta t$  for the tangential velocities $v=0.3$ (a) and 0.6 m/s (b). Curves and error bars are averages and standard deviations estimated from 10 experiments with particles distributed uniformly along their circular trajectory.  Solid and dashed curves correspond to trajectory radii of 50 and 100 mm.}
    \label{fig:s vs dt}
\end{figure}

Next, we run multiple MCMCs with increasing $\Delta t$ in order to assess the trade-off between (i) the reduction in uncertainty associated with the larger number of LoRs, and (ii) the increase in uncertainty caused by the particle displacement during  $\Delta t$. Indeed, in the absence of motion, we expect $s$ to decrease as $N^{-1/2} \approx (\rho\Delta t)^{-1/2}$, but the particle displacement introduces an error in the model underpinning \eqref{eq:probabilityfull} that increases with $\Delta t$. 
One may wonder why the activity $\rho$ is not increased to alleviate the uncertainty in the moving particle position. Increasing $\rho$ provides more data over the same $\Delta t$ and hence monotonically decreases the uncertainty. In fact, $\rho$ may be sufficiently increased such that the particles are quasi-static over $\Delta t$, in which case $s\propto \rho^{-1/2}$. However, in PEPT, as opposed to PET, activity cannot be increased by simply introducing more radioactive material with the same activity density. Here each particle is radiolabeled separately, and the requirement that particles remain small so that they are easily suspended in fluid makes it challenging to increase each particle activity. We therefore analyze particles with fixed activity, rotating with the two tangential velocities 0.3 and 0.6 m/s obtained by changing either the frequency or radius of the circular trajectory. In this way, we can examine the effect of varying the acceleration while maintaining a constant velocity. The results are shown in  figures \ref{fig:s vs dt T0 v0.3} for $v=0.3$ m/s and \ref{fig:s vs dt T0 v0.6} for $v=0.6$ m/s. 
The blue curves, with the left vertical axis, show the metric $s$, averaged over the 10 positions of the particle, as a function of $\Delta t$, with the error bars indicating the corresponding standard deviation. The red curves, with the right vertical axis, show the absolute error $\left|\bm{x}-\bm{x}_*\right|$ between the average inferred position $\bm{x}$ and the particle actual position $\bm{x}_*$. The solid and dashed curves are calculations for particles rotating with $r=50$ and $100$, respectively, corresponding to a twofold decrease in particle acceleration. The scales in both the left and right vertical axes in figure \ref{fig:s vs dt} are equal for ease of comparison.

All the uncertainty curves (blue, left vertical axis) exhibit minima, reflecting the trade-off mentioned above. As could be expected, $s$ is smaller for $v=0.3$ m/s  than for $v=0.6$ m/s as a result of smaller particle displacements with the same rate of LoR generation, and for the same velocity, $s$ is smaller for the smaller acceleration.
The curves are asymmetric about the minima, with a sharp decrease, and a much milder increase (and even some further decrease at large $\Delta t$ for $v=0.6$ m/s and $r=100$ mm (solid curve in figure \ref{fig:s vs dt T0 v0.6})). The error (red, right vertical axis), in contrast, increases unambiguously for large  $\Delta t$, reflecting the expected model's inability to accurately infer the instantaneous position when the particle moves significantly over $\Delta t$. To provide reference to our results, we also analyzed the data with the Birmingham algorithm \cite{Parker1993}, the most commonly used method for PEPT inversion \cite{Windows-Yule2022}, and the expectation-maximization algorithm \cite{Manger2021}. The error $\left|\bm{x}-\bm{x}_*\right|$ achieved by both methods (not shown in the figures) is roughly 1 mm -- typically larger than our error at small $\Delta t$ but smaller at large $\Delta t$. Other inversion algorithms may be tested and compared against our results; however, our primary aim is to quantify the uncertainty in PEPT rather than developping an alternative algorithm. Accordingly, no attempts were made to optimise our method by, for instance, testing various priors.

\begin{figure}
\centering
\subfloat[][]{\includegraphics[trim=0 5mm 0 0,clip,scale=0.272]{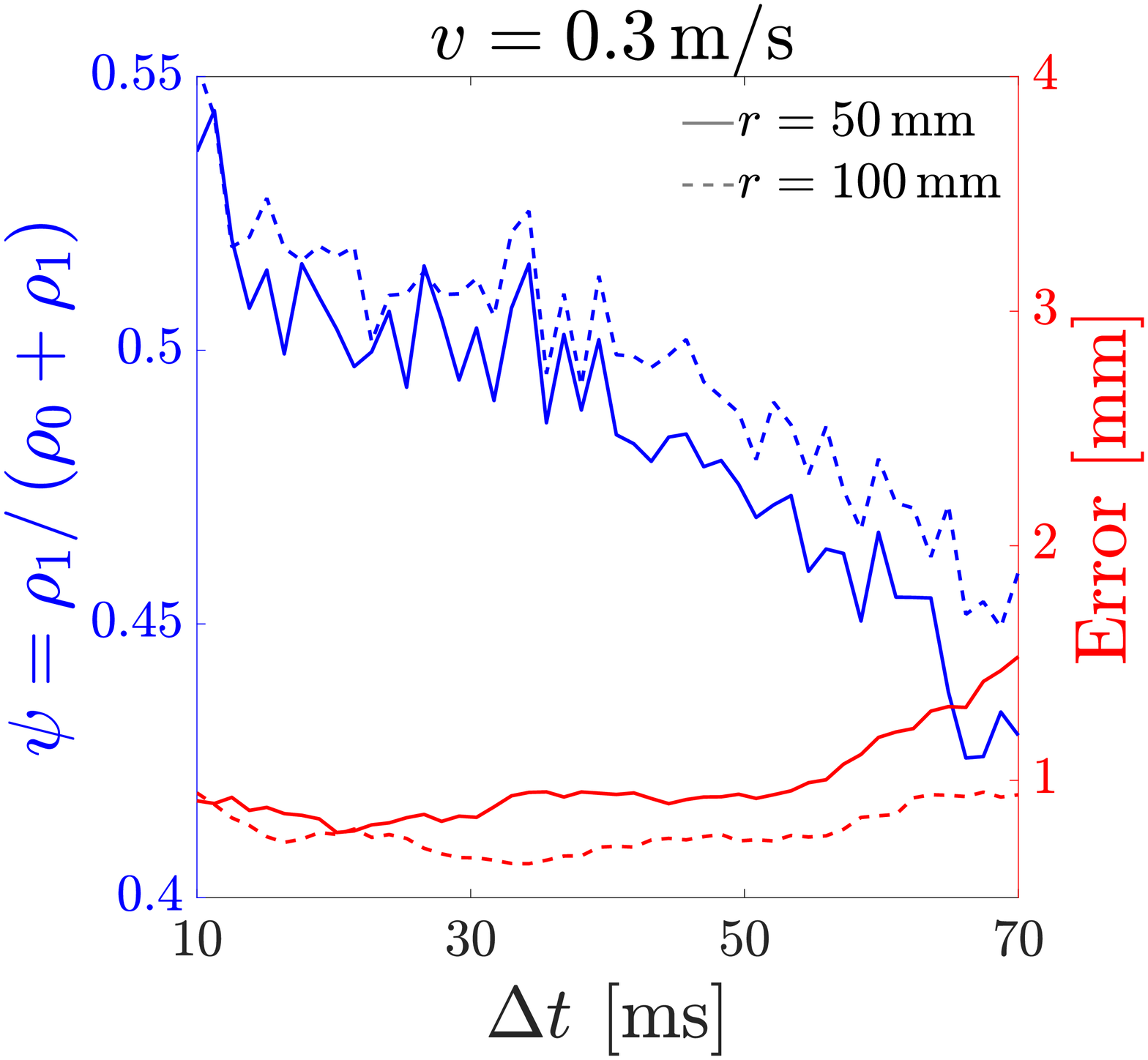}
\label{fig:rho and error v03}}
\qquad
\subfloat[][]{\includegraphics[trim=0 4.5mm 0 0,clip,scale=0.27]{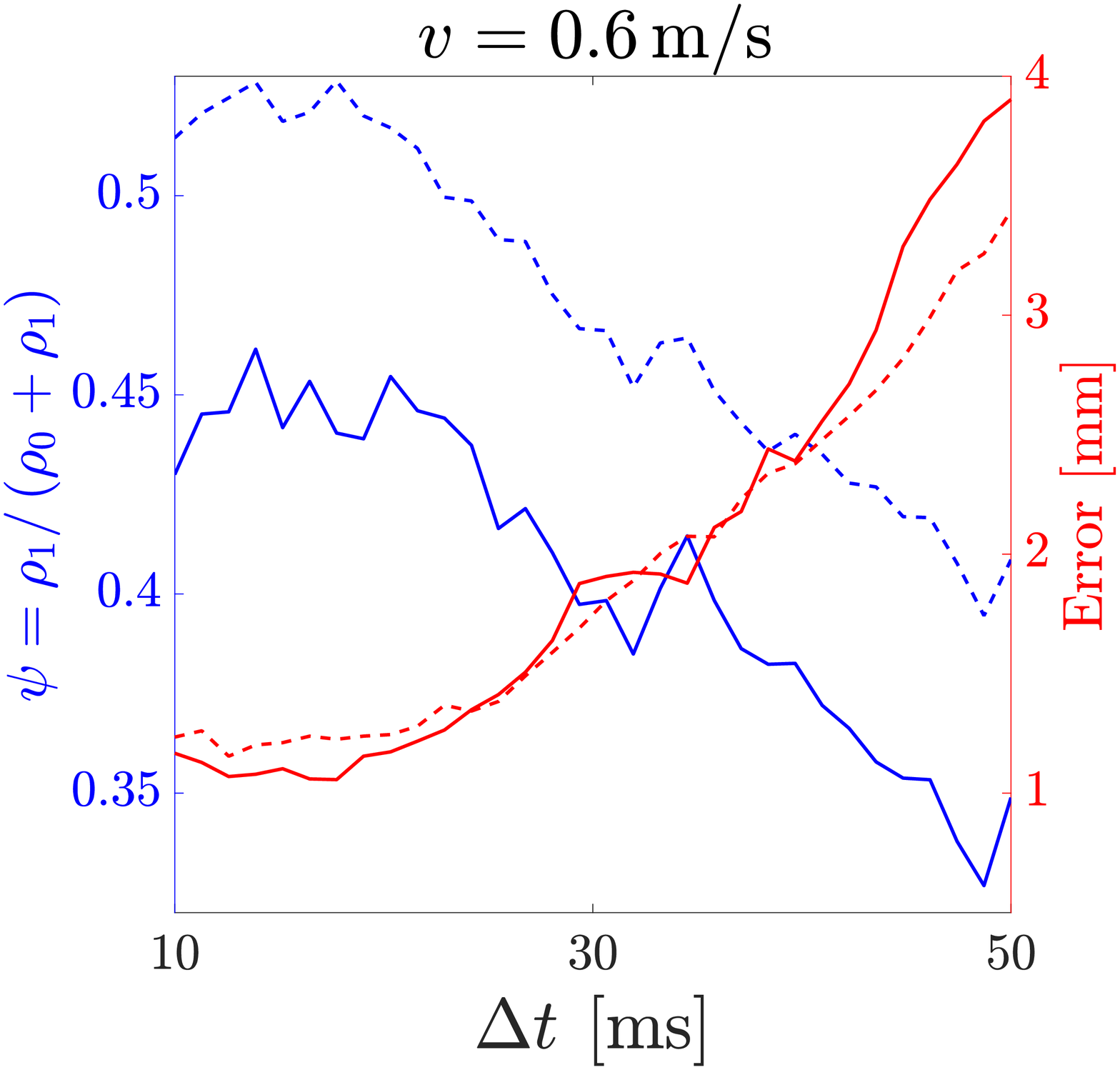}
\label{fig:rho and error v06}}

    \caption{Inferred ratio $\psi=\rho_1/(\rho_0+\rho_1)$ of rate of unscattered LoR detection to total LoR detections (blue curves, left axes) and error in inferred and error $|\bm{x}-\bm{x_*}|$ in the inferred  instantaneous particle position (red curves, right axes) as functions of the observation interval $\Delta t$  for the tangential velocities $v=0.3$ (a) and 0.6 m/s (b). Curves are averages from 10 experiments, and  solid and dashed curves correspond to trajectory radii of 50 and 100 mm, as in figure \ref{fig:s vs dt}.}
    
    \label{fig:activity and error}
\end{figure}

It might seem paradoxical that the uncertainty does not increase substantially with increasing $\Delta t$ while the error does. This is the result of our weakly constrained model of scattering: LoRs that are inconsistent with emission by a fixed particle are attributed to the scattered component of the likelihood, whether they arise because of scattering or because of particle motion. This explanation is confirmed in figure \ref{fig:activity and error} which shows the inferred relative activity  $\psi=\rho_1/\left(\rho_0+\rho_1\right)$ of unscattered LoRs (blue curves, left axis). Although, the correct, physical activity should be independent of $\Delta t$, the inferred value decreases, as a result of the increasing misattribution of LoRs to the scattered component.  
One way of remedying this artifact  is to constrain the ratio $\rho_1/\rho_0$ or, equivalently, $\psi$ to a physically sensible range using the prior. Another way is to account for the particle motion in the likelihood. We adopt the latter approach next by implementing the Taylor-expansion-based first-order inference described in \S\ref{sec:bayesian}.

\subsection{First-order inference} \label{sec:high-order inference}

The first-order inference relies on the form \eqref{eq:Bayes_t} of Bayes' formula, with the time-dependent likelihood \eqref{eq:probability PoissonTime} and a linear approximation to the particle trajectory, so both the particle position and velocity in the middle of the observation interval need to be inferred. For simplicity, we approximate $\bm{x}(t) \approx \bm{x}(t_0) = \mathrm{const.}$ in the factor $\Lambda$ in \eqref{eq:Lambda0} and use expression \eqref{eq:Lambda} that depends on positions only. The dependence on velocities is then through the rates $\lambda(L|\bm{x}_k(t),\rho_k)$.

\begin{figure}
\centering
\subfloat[][]{\includegraphics[trim=0 5mm 0 0,clip,scale=0.26]{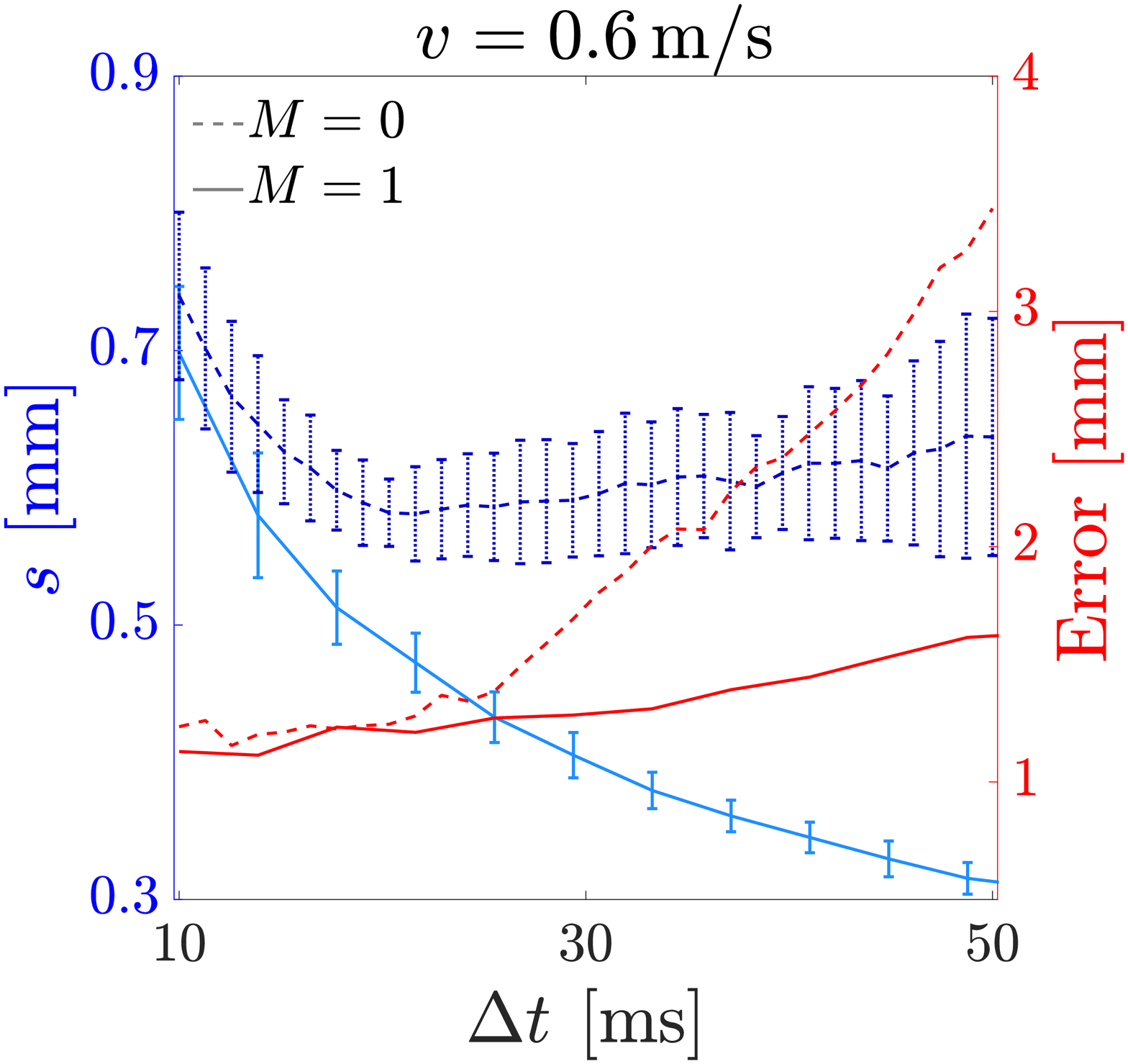}
\label{fig:1st order position}}
\qquad
\subfloat[][]{\includegraphics[trim=0 14mm 0 0,clip,scale=0.282]{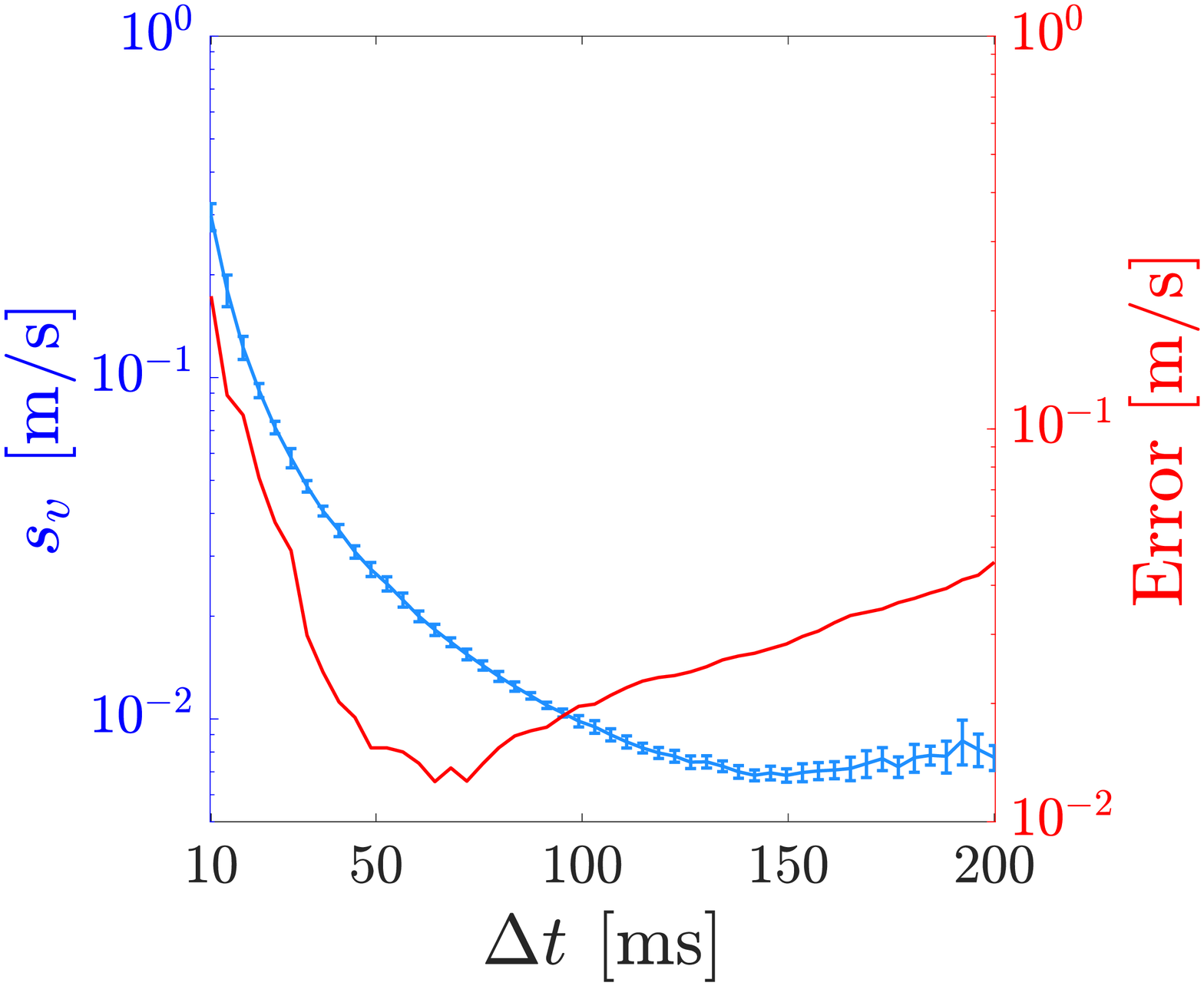}
\label{fig:1st order velocity}}

    \caption{(a) Comparison between inference of activity and position ($M=0$, dashed curves) and inference of activity, position and velocity ($M=1$, solid curves) for a particle on a circular trajectory with $r=100$ mm and $v=0.6$ m/s: uncertainty metric $s$ (blue curves, left axis) and absolute error $|\bm{x}-\bm{x}_*|$ (red curves, right axis).   (b) Uncertainty $s_v$ (blue curve, left axis) and error $|\bm{v}-\bm{v}_*|$ on the velocity (red curve, right axis) inferred for $M=1$. In both (a) and (b), curves and error bars are averages and standard deviations estimated from 10 experiments with particles distributed uniformly along their circular trajectory.} 
    \label{fig:1st order tracking}
\end{figure}

We carry out MCMC sampling for the first-order ($M=1$) inference in the case $r=100$ mm and $v=0.6$ m/s and compare it with the results of the position-only inference ($M=0$) of the previous section. For this comparison we use the same number of MCMC steps for $M=1$ as for $M=0$, even though the number of parameters to estimate increases from $5$ to $8$. The results are shown in figure \ref{fig:1st order position} which shows the uncertainty metric $s$ (blue, left axis) and the position error (red, right axis), with the same conventions as in figure \ref{fig:s vs dt}.

Both the $M=1$ uncertainty and error curves lay below their counterparts for $M=0$. This demonstrates that inferring the instantaneous velocity increases the measurement accuracy.
Furthermore, the variance of the set of 10 experiments, indicated by the error bars, is also smaller. The difference between the errors for $M=0$ and 1 is marginal at $\Delta t \lesssim 20$ ms, where the particle moves less than $v\Delta t \approx 12$ mm over the time interval. At larger $\Delta t$, however, the $M=0$ error curve rises at a much steeper rate compared with the $M=1$ one. The increased particle movement is then tackled better by inferring the particle trajectory over $\Delta t$ as linear in $t$ rather than fixed. Note that we carried out the $M=1$ inference for $\Delta t$ up to $200$ ms (not shown in figure \ref{fig:1st order position}) to verify that $s$ reaches a minimum value, found to be $\Delta t\approx150$ ms. 

A benefit of $M=1$ inference over $M=0$, in addition to an increase accuracy, is that it estimates directly the particle velocity, often the main quantity of interest for fluid dynamical applications, which otherwise needs to be deduced from the positions, e.g., using finite differences. We show the uncertainty and error on the inferred particle velocity in figure  \ref{fig:1st order velocity}. The blue curve (left axis) shows the mean (over 10 particle positions, as before) uncertainty metric $s_v$ defined as in \eqref{eq:s} but using the eigenvalues of the velocity covariance matrix, with error bars indicating the standard deviation. The red curve shows the mean error $|\bm{v}-\bm{v}_*|$ between the inferred particle velocity $\bm{v}$ and the exact velocity of the circular motion $\bm{v}_*$. The qualitative behavior of the uncertainty and error on the inferred velocity is roughly similar to that corresponding to the position, though the minimum error appears for a larger $\Delta t$. The decrease in inferred uncertainty as $\Delta t$ increases is related to the misattribution of LoRs to the scattered components, as discussed in \S\ref{sec:results}.

\section{Conclusion}

In this paper, we propose a probabilistic model for the detection of random LoRs emitted by moving positron-emitting particles with prescribed trajectories and activities. 
The model lays the foundation for the systematic quantification of uncertainty in PEPT. For simplicity, the model bundles various sources of uncertainty into a single Gaussian error for the position of the LoRs.  It would however be straightforwardly generalized to include detailed representations of specific uncertainty sources including positron range, discrete detectors, imperfect alignment of the paths of the pair of photons, etc.\ 

The generation of LoRs is described as a Poisson process in the space of LoRs, which leads to a general formulation largely independent of the details of the PEPT device, in particular of the detector configuration. We give an explicit formula for the Poisson rate in the case of a cylindrical configuration. The derivation involves various geometric properties and approximations. It incorporates a model of scattering that assumes that scattering occurs at a rate independent of the position of the radioactive particles and leads to a uniform distribution of LoRs. These are major simplifications whose relaxation would introduce technical rather than conceptual complications. 

The likelihood associated with the Poisson process makes it possible to use a Bayesian approach to infer the position and activity of the radioactive particles. This has benefits over existing PEPT inversion methods including the systematic nature of the inference, its capability to include the model refinements mentioned above, and the quantification of uncertainty. Furthermore, the Bayesian formulation is directly adapted to the inference of particle trajectories and time-dependent activities instead of fixed positions and activities. 
We take advantage of this to propose a hierarchy of Bayesian inversion procedures, based on a $M$-term Taylor expansion of the particle trajectories, which requires the inference of the $M$ first derivatives of the positions in addition to the positions themselves. We demonstrate the potential of these procedures by carrying out a full Bayesian uncertainty quantification for a single rotating particle using $M=0$ and $M=1$ inference. The Taylor-expansion-based procedures remain local in time in that they are carried out independently in small time interval. It will be of interest to develop alternative Bayesian procedures that infer entire trajectories, parameterized, for instance, as splines (cf.\ \cite{Lee2015}). We leave this for future work.

\section*{Acknowledgments}
This research was supported by EPSRC Programme Grant EP/R045046/1: Probing Multiscale Complex Multiphase Flows with Positrons for Engineering and Biomedical Applications (PI: Prof. M. Barigou, University of Birmingham).

\appendix
\section{Mapping $F$ and covariance matrix $\Sigma$} \label{ap:mapping}

We first derive an explicit form for the mapping $F$ between the representation
\begin{equation}
L=\left(\varphi,\theta,a_{\varphi},a_{\theta}\right)\in\left[0,2\pi\right]\times\left[0,\pi/2\right]\times\mathbb{R}^{2}
\end{equation}
of LoRs and the coordinates
\begin{equation}
u=\left(\phi_{1},\phi_{2},z_{1},z_{2}\right)\in\left[0,2\pi\right]\times\left[0,2\pi\right]\times\left[-H/2,H/2\right]\times\left[-H/2,H/2\right]
\end{equation}
of their two intersections with the cylindrical detector. A parametric representation of an LoR is $\bm{x} = a_{\varphi}\bm{e}_{\varphi}+a_{\theta}\bm{e}_{\theta}+ \omega \bm{e}_r $, with $\omega \in \mathbb{R}$ the parameter such that $\omega=0$ corresponds to the point closest to the origin. Using standard expressions for the $\bm{e}_\varphi, \, \bm{e}_\theta$ and $\bm{e}_r$ in terms of the  canonical Cartesian basis, we can write the condition that the LoR intersects the cylinder of radius $R$ at $(\phi,z)$ as
\begin{align}
a_{\theta}\cos\theta\cos\varphi-a_{\varphi}\sin\varphi+\omega\sin\theta\cos\varphi&=R\cos\phi, \label{eq:mapping x} \\
a_{\theta}\cos\theta\sin\varphi+a_{\varphi}\cos\varphi+\omega\sin\theta\sin\varphi&=R\sin\phi, \label{eq:mapping y} \\
-a_{\theta}\sin\theta+\omega\cos\theta&=z. \label{eq:mapping z}
\end{align}
Adding the squares of \eqref{eq:mapping x} and \eqref{eq:mapping y} gives the values
\begin{equation}
\omega_{1,2}=\frac{-a_{\theta}\cos\theta\pm\Upsilon}{\sin\theta}, \label{eq:xr}
\end{equation}
where $\Upsilon =\sqrt{R^2-a_{\varphi}^2}$,
for the parameter corresponding to each intersection.
Substituting these values into \eqref{eq:mapping x}--\eqref{eq:mapping z} gives the mapping  $u=F\left(L\right)$ as
\begin{equation}
\left(\begin{array}{c}
\phi_{1}\\
\phi_{2}\\
z_{1}\\
z_{2}
\end{array}\right)=\left(\begin{array}{c}
\tan^{-1} \left(\left(a_{\varphi}\cos\varphi+\Upsilon\sin\varphi\right)/\left(-a_{\varphi}\sin\varphi+\Upsilon\cos\varphi\right)\right)\\
\tan^{-1}\left(\left(-a_{\varphi}\cos\varphi+\Upsilon\sin\varphi \right)/\left(a_{\varphi}\sin\varphi+\Upsilon\cos\varphi\right)\right)\\
{(-a_{\theta}+\Upsilon\cos\theta)}/{\sin\theta}\\
{(-a_{\theta}-\Upsilon\cos\theta)}/{\sin\theta}
\end{array}\right). \label{eq:F mapping}
\end{equation}

We next compute the covariance $\Sigma$ from \eqref{L|L'}. We calculate the Jacobian matrix $\nabla F$ and invert it to obtain
\begin{align}
&\left(\nabla F\left[u\right]\right)^{-1}=\frac{1}{2\Upsilon} \times \\ 
&\left(\begin{array}{cccc}
\Upsilon & \Upsilon & 0 & 0\\
-a_{\varphi}\cos\theta\sin\theta & a_{\varphi}\cos\theta\sin\theta & -\sin^{2}\theta & \sin^{2}\theta\\
\Upsilon^{2} & -\Upsilon^{2} & 0 & 0\\
-a_{\varphi}x_{\theta}\cos^{2}\theta & a_{\varphi}x_{\theta}\cos^{2}\theta & -\sin\theta\left(\Upsilon+a_{\theta}\cos\theta\right) & \sin\theta\left(\Upsilon-a_{\theta}\cos\theta\right)
\end{array}\right). \nonumber
\end{align}
The covariance matrix then follows from \eqref{L|L'} as
\begin{align} & \Sigma=\frac{g^{2}}{2R^{2}} \times \label{eq:Sigma}\\
 & \left(\begin{array}{cccc}
1 & 0 & 0 & 0\\
 & \sin^{2}\theta\left(R^{2}/\Upsilon^{2}\left(1+\sin^{2}\theta\right)-\cos^{2}\theta\right)/2 & -a_{\varphi}\sin\left(2\theta\right)/2 & a_{\theta}\sin\left(2\theta\right)\left(R^{2}/\Upsilon^{2}-\sin^{2}\theta\right)/2\\
  &  & \Upsilon^{2} & -a_{\varphi}a_{\theta}\cos^{2}\theta\\
 &  &  & R^{2}\sin^{2}\theta+a_{\theta}^{2}\cos^{2}\theta\left(R^{2}/\Upsilon^{2}-\cos^{2}\theta\right)
\end{array}\right),\nonumber
\end{align}
where we use that $\Sigma=\Sigma^{\mathrm{T}}$ to avoid  writing the lower triangle entries.
The final covariance matrix $\Sigma_2$ is simply the $2\times2$ lower diagonal block of \eqref{eq:Sigma}, that is 
\begin{equation}
\Sigma_{2}=\frac{g^{2}}{2R^{2}}\left(\begin{array}{cc}
\Upsilon^{2} & -a_{\varphi}a_{\theta}\cos^{2}\theta\\
-a_{\varphi}a_{\theta}\cos^{2}\theta & R^{2}\sin^{2}\theta+a_{\theta}^{2}\cos^{2}\theta\left(R^{2}/\Upsilon^{2}-\cos^{2}\theta\right)
\end{array}\right). \label{eq:Sigma2}
\end{equation}

\section{Derivation of $\sigma_{0}^{2}$}\label{ap:sigma0}

We evaluate the outlier variance \eqref{eq:sigma0 def}, defined as the volume of detectable LoRs:
\begin{equation}
\sigma_{0}^{2} = \int_\mathcal{D} \d\mu=\int_{0}^{2\pi}\d\varphi\int_{-R}^{R}\d a_{\varphi}\int_{\overline{\theta}_{\text{min}}}^{\pi/2}\d\theta\int_{a_{\theta}^{-}}^{a_{\theta}^{+}}\d a_{\theta}. \label{eq:sigma0 expanded}
\end{equation}
Here, the integration limits $\overline{\theta}_{\text{min}}$ and $a_\theta^\pm$ are determined by the condition that the LoR intersects the cylindrical detector twice. They are found by first substituting \eqref{eq:xr} into \eqref{eq:mapping z} to obtain the heights of the two intersections points
\begin{equation}
z_{1,2}=\frac{-a_{\theta}\pm\Upsilon\cos\theta}{\sin\theta},
\end{equation}
then requiring that $-H/2\leq z_{1,2}\leq H/2$, to find
\begin{equation}
a_{\theta}^{\pm} =\mp \Upsilon\cos\theta \pm {H\sin\theta}/{2}. \label{eq:ythetamin}
\end{equation}
Equating these gives the minimum angle for detectable LoRs,
\begin{equation}
\overline{\theta}_{\text{min}}=\tan^{-1} \left(2\Upsilon/{H}\right). \label{eq:thetamin bar}
\end{equation}
Integrating \eqref{eq:sigma0 expanded} using \eqref{eq:ythetamin}--\eqref{eq:thetamin bar} gives expression \eqref{eq:sigma0^2} for $\sigma_0^2$.

\begin{figure}
    \centering
    \includegraphics[scale=0.1]{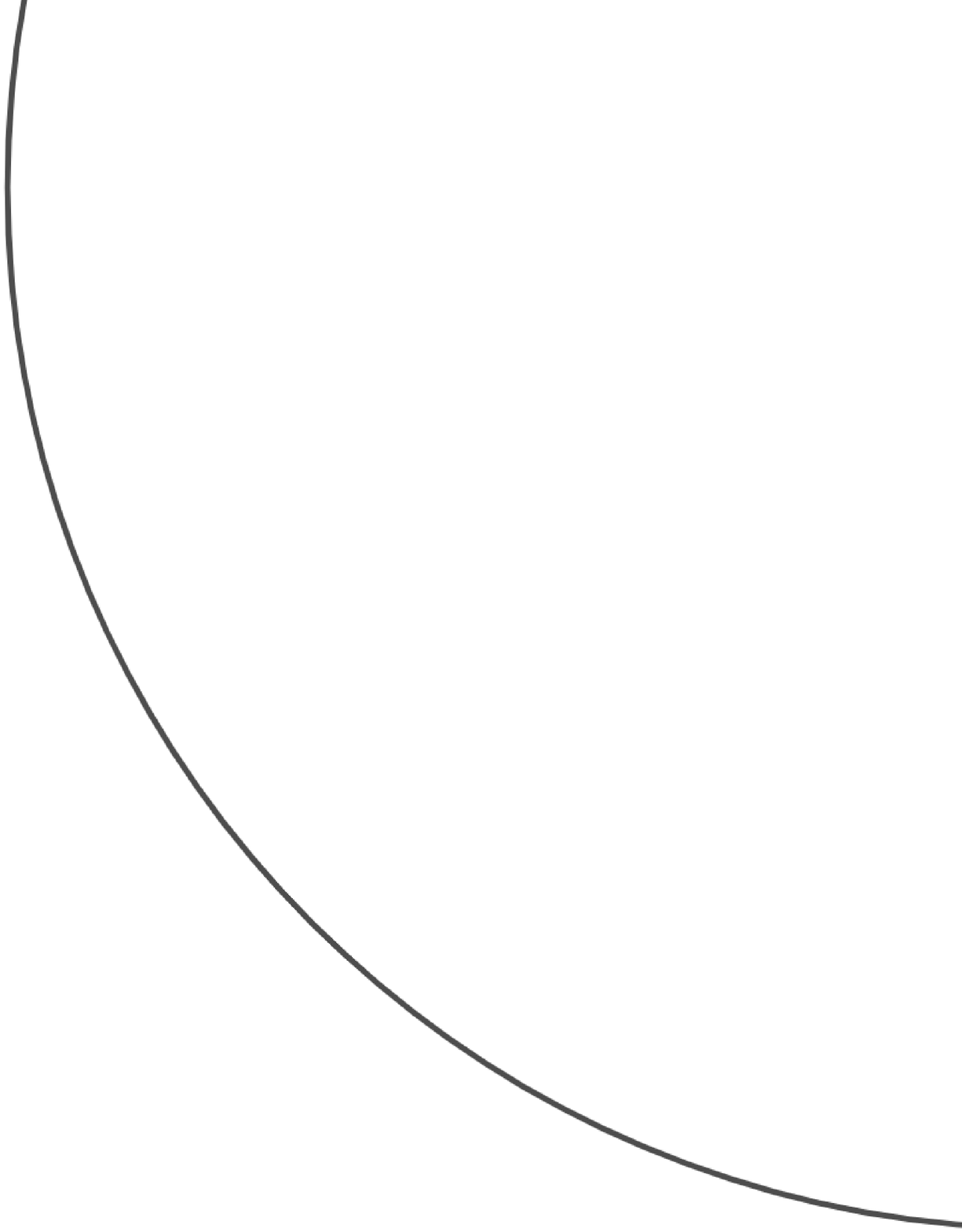}
    \caption{Top view of a cylindrical detector surface, with a representative LoR (red) emanating from a particle at $\bm{x}=\left\{r,z\right\}$ (black dot).}
    \label{fig:circle}
\end{figure}

\section{Derivation of $\mathcal{G}$} \label{ap:G}

We obtain an explicit form for $\mathcal{G}(\bm{x})$ in \eqref{eq:G_k def} which can be interpreted as a solid angle subtended by the cylindrical detector from the point $\bm{x}$. By symmetry, this depends only on the cylindrical coordinates $(r,z)$ of the point $\bm{x}$ which we can assume to be along the $x$-axis. The geometry of the LoR is then as depicted 
in figure \ref{fig:circle} when  projected on the plane through the origin perpendicular to the generator of the cylinder. The lengths $b_\pm$ of the projections of the segments of the LoR above and beyond the planes are  found as

\begin{equation}
b_{\pm}=\sqrt{R^{2}-\left(r\sin\varphi\right)^{2}}\pm r\cos\varphi.
\end{equation}
The conditions for the LoR to intersect twice the detector are then that $b_\pm \cot \theta \le  z \pm H/2$, hence the minimal value of $\theta$ for two intersections is
\begin{equation}
\theta_{\text{min}}\left(r,z,\varphi;R,H\right)=\max\left[\tan^{-1}\left(\frac{b_{-}}{H/2-z}\right),\tan^{-1}\left(\frac{b_{+}}{H/2+z}\right)\right].
\label{eq:thetamin}
\end{equation}

It is convenient to non-dimensionalize variables by letting $r=R\hat{r}\,,\,z=H\hat{z}$, and $\beta=R/H$. Eq.\ \eqref{eq:thetamin} can then be written explicitly as
\begin{equation}
\theta_{\text{min}}\left(\hat{r},\hat{z},\varphi;\beta\right)=\begin{cases}
\tan^{-1}\left(\beta f_{-}\right) & \textrm{for} \  \hat{r}<\left|2\hat{z}\right|, \ \textrm{and for} \ \hat r > \left|2\hat{z}\right| \ \textrm{and} \  \varphi>\alpha,\\
\tan^{-1}\left(\beta f_{+}\right) & \textrm{for} \ \hat{r}> \left|2\hat{z}\right| \ \textrm{and} \  \varphi<\alpha,
\end{cases} \label{eq:ThetaMin}
\end{equation}
where we have defined
\begin{equation}
f_{\pm}\left(\hat{r},\hat{z},\varphi\right)=\frac{\sqrt{1-\left(\hat{r}\sin\varphi\right)^{2}}\pm\hat{r}\cos\varphi}{1/2\pm \hat{z}}, \
\alpha\left(\hat{r},\hat{z}\right)=\cos^{-1} \sqrt{\frac{\left(2\hat{z}\right)^{2}}{1-\left(2\hat{z}\right)^{2}}\frac{1-\hat{r}^{2}}{\hat{r}^{2}}}.
\end{equation}

Substituting \eqref{eq:ThetaMin} into \eqref{eq:G_k def} yields
\begin{equation}
\mathcal{G}\left(\hat{r},\hat{z}\right)=\begin{cases}
\displaystyle{\frac{1}{\pi} \int_{0}^{\pi} (1+\beta^{2}f_{-}^{2})^{-1/2} \, \d\varphi} & \textrm{for} \  \hat{r}<\left|2\hat{z}\right|,\\
 \displaystyle{\frac{1}{\pi} \left(\int_{0}^{\alpha} (1+\beta^{2}f_{+}^{2})^{-1/2} \, \d\varphi+ \int_{\alpha}^{\pi} (1+\beta^{2}f_{-}^{2})^{-1/2} \, \d\varphi \right)} & \textrm{for} \ \hat{r}>\left|2\hat{z}\right|.
\end{cases} \label{eq:Gk appendix}
\end{equation}
The integrals can be evaluated numerically.

In the limits $\beta\ll1$ and $\beta\gg1$ corresponding to a narrow vertical cylinder and a ring, respectively, the integrands may be simplified using a series expansion in $\beta$ and $\beta^{-1}$
\begin{equation}
\frac{1}{\sqrt{1+\beta^{2}f_{\pm}^{2}}}=\begin{cases}
1+\sum_{n}\left(-1\right)^{n}c_{n}\beta^{2n}f_{\pm}^{2n} & \textrm{for} \ \beta\ll1,\\
\sum_{n}\left(-1\right)^{n+1}c_{n}\beta^{1-2n}f_{\pm}^{1-2n}, & \textrm{for} \ \beta\gg1.
\end{cases}, \label{eq:series}
\end{equation}
with
$c_{n}={1}/{2}, \, {3}/{8}, \, {5}/{16},\, {35}/{128}, \, {63}/{256},\cdots
$
Substituting \eqref{eq:series} into \eqref{eq:Gk appendix}, we  obtain
\begin{align}
\mathcal{G}\left(\hat{r},\hat{z};\beta\right)&=1-\frac{2\beta^{2}}{\left(1-2\hat{z}\right)^{2}}+\frac{4\beta^{2}\mathcal{H}\left(\hat{r}-\left|2\hat{z}\right|\right)}{\pi\left(1-4\hat{z}^{2}\right)^{2}}\Biggl[2\hat{z}\left(2\alpha+\hat{r}^{2}\sin2\alpha\right)-\left(1+4\hat{z}^{2}\right)\times \label{eq:Gk beta small} \\ 
&\left(\arcsin\left(\hat{r}\sin\alpha\right)+\hat{r}\sin\alpha\sqrt{1-\left(\hat{r}\sin\alpha\right)^{2}}\right)\Biggr]+O\left(\beta^{4}\right),\quad \textrm{for} \  \beta\ll1 \nonumber \\
\mathcal{G}\left(\hat{r},\hat{z};\beta\right)&=\frac{\left(1-2\hat{z}\right)}{6\pi\beta\left(1-\hat{r}^{2}\right)}\Biggl(F\left(\hat{r}^2\right)\left[\frac{\left(1-2\hat{z}\right)^{2}}{\beta^{2}}+\mathcal{H}\left(\left|2\hat{z}\right|-\hat{r}\right)\frac{\hat{r}^{2}\left(4+7\hat{r}^{2}-3\hat{r}^{4}\right)}{\left(1-\hat{r}^{2}\right)}\right] \label{eq:Gk beta large} \\
&+6E\left(\hat{r}^{2}\right)\left[1-\frac{\left(1-2\hat{z}\right)^{2}\left(7+\hat{r}^{2}\right)}{24\beta^{2}\left(1-\hat{r}^{2}\right)^{2}}\right]\Biggr)+\frac{\mathcal{H}\left(\hat{r}-\left|2\hat{z}\right|\right)}{\pi\beta\left(1-\hat{r}^{2}\right)}\times \nonumber \\
&\Biggl(\left[2\hat{z}E\left(\alpha | \hat{r}^{2}\right)\left(1-\frac{\left(7+\hat{r}^{2}\right)\left(3+4\hat{z}^{2}\right)}{24\beta^{2}\left(1-\hat{r}^{2}\right)^{2}}\right)-\hat{r}\sin\alpha\right]+\frac{1}{6\beta^{2}\left(1-\hat{r}^{2}\right)^{2}}\times \nonumber \\
&\biggl[\frac{\hat{r}\left(1+12\hat{z}^{2}\right)\left(9\sin\alpha+\hat{r}^{2}\sin3\alpha\right)}{4}+\hat{z}\left(3+4\hat{z}^{2}\right)\times \nonumber \\
&\left(2F\left(\alpha|\hat{r}^{2}\right)\left(1-\hat{r}^{2}\right)-\hat{r}^{2}\sin2\alpha\sqrt{1-\left(\hat{r}\sin\alpha\right)^{2}}\right)\biggr]\Biggr)+O\left(\beta^{-5}\right),\quad \textrm{for} \ \beta\gg1 \nonumber
\end{align}
where $\mathcal{H}$ is the Heaviside step function, and $F$ and $E$ are the incomplete elliptic integrals of the first and second kind, respectively,
\begin{equation}
F\left(\left.\Theta\right|\xi\right) =\int_{0}^{\Theta}\left(1-\xi\sin^{2}\varphi\right)^{-1/2}\,\d\varphi, \quad
E\left(\left.\Theta\right|\xi\right)=\int_{0}^{\Theta}\left(1-\xi\sin^{2}\varphi\right)^{1/2}\,\d\varphi,
\end{equation}
with $F\left(\xi\right)= F\left(\pi/2,\xi\right)$ and $E\left(\xi\right)= E\left(\pi/2,\xi\right)$ the corresponding complete integrals.

\begin{figure}
\centering
\subfloat[][]{\includegraphics[trim=0 0 1.5cm 0,clip,scale=0.27]{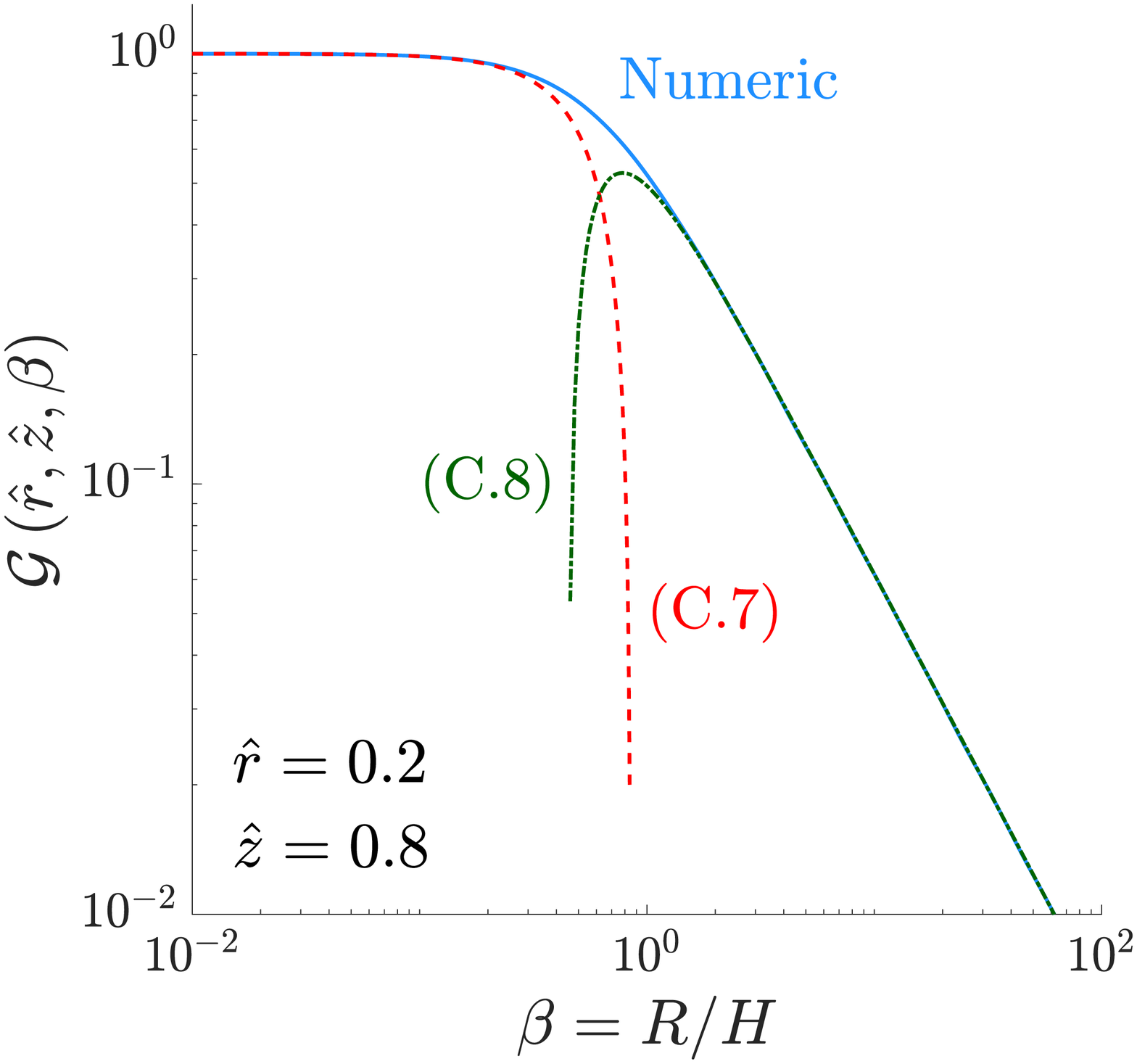}
\label{fig:G r small}}
\qquad
\subfloat[][]{\includegraphics[trim=0 0 1.5cm 0,clip,scale=0.27]{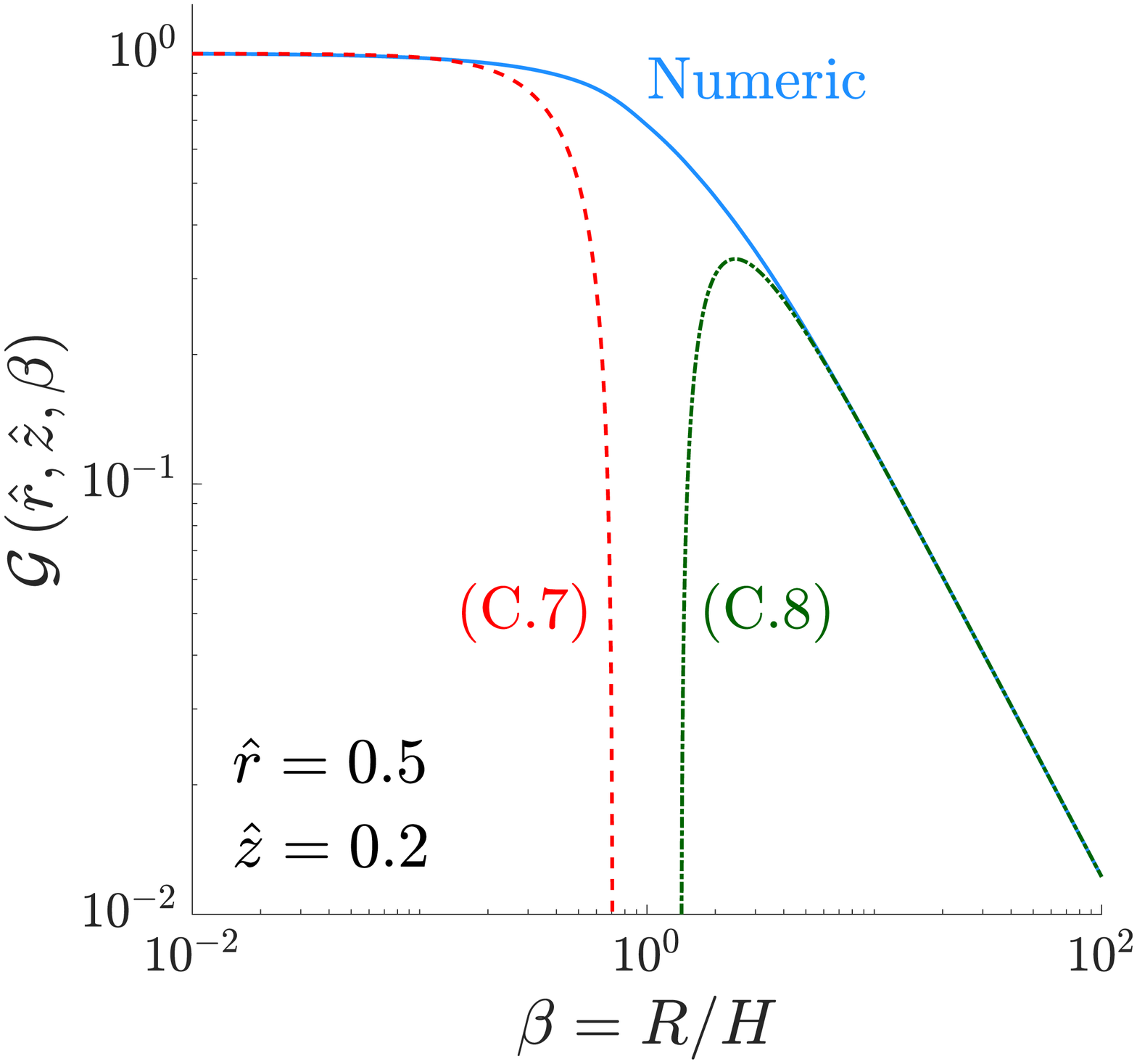}
\label{fig:G r large}}

    \caption{The geometric function $\mathcal{G}$, calculated for $\hat{r}<\left|2\hat{z}\right|$ (a) and$\hat{r}>\left|2\hat{z}\right|$ (b) as a function of the detecting surface aspect ratio $\beta=R/H$. The solid blue curves are the numerical calculation of \eqref{eq:Gk appendix}; the dashed red and dashed-dotted green curves are the asymptotic solutions \eqref{eq:Gk beta small} and \eqref{eq:Gk beta large}.}
    \label{fig:G}
\end{figure}


\end{document}